%% file: paper.tex
\documentclass[sigconf, screen]{acmart}
\renewcommand\footnotetextcopyrightpermission[1]{} 

\copyrightyear{2020}
\acmYear{2020}
\setcopyright{acmlicensed}
\acmConference[CIKM '20]{Proceedings of the 29th ACM International Conference on Information and Knowledge Management}{October 19--23, 2020}{Virtual Event, Ireland}
\acmBooktitle{Proceedings of the 29th ACM International Conference on Information and Knowledge Management (CIKM '20), October 19--23, 2020, Virtual Event, Ireland}
\acmPrice{15.00}
\acmDOI{10.1145/3340531.3412759}
\acmISBN{978-1-4503-6859-9/20/10}

\usepackage{glossaries}
\usepackage{natbib}
\usepackage{pifont} 
\usepackage{multirow}  
\usepackage{subcaption} 
\usepackage[font={footnotesize}]{caption} 
\usepackage{placeins}
\usepackage[inline]{enumitem}
\usepackage[noend, ruled, linesnumbered, commentsnumbered]{algorithm2e}
\usepackage{adjustbox}
\usepackage{color, colortbl}
\definecolor{myBlue}{RGB}{55, 126, 184}
\definecolor{myGreen}{RGB}{77, 175, 74}
\definecolor{Gray}{gray}{0.9}
\newcommand{\para}[1]{\smallskip\noindent{\bf{#1}}}

\loadglsentries{glossaries.tex}

\title[MindReader: Recommendation over Knowledge Graph Entities]{MindReader: Recommendation over Knowledge Graph Entities \\ with Explicit User Ratings}

\author{Anders H. Brams, Anders L. Jakobsen,
Theis E. Jendal, Matteo Lissandrini, 
Peter Dolog, Katja Hose}
\email{{ahbr, alja, tjendal, matteo, dolog, khose}@cs.aau.dk}
\affiliation{%
    \department{Department of Computer Science}
    \institution{Aalborg University}
}

\begin{document}
\fancyhead{}
\input{rules}
\input{commands}

\begin{abstract}
\input{sections/abstract}
\end{abstract}

\maketitle
\pagestyle{plain} 
\input{sections/introduction}
\input{sections/motivation}
\input{sections/related}


\input{sections/methodology}
\input{sections/analysis}
\input{sections/evaluation}

\input{sections/experiments}
\input{sections/results}
\input{sections/discussion}
\input{sections/conclusions}

\para{Acknowledgements.}
Matteo Lissandrini is supported by the European Union’s Horizon 2020 Research and Innovation Programme under the Marie Skłodowska-Curie grant agreement no. 838216. Katja Hose and Theis Jendal are supported by the Poul Due Jensen Foundation. We want to thank the MindReader users for their help in building the dataset.

\bibliography{bibliography_short}
\clearpage
\appendix
\input{sections/appendix}

\end{document}

%% file: glossaries.tex
\newacronym{kg}{KG}{Knowledge Graph}
\newacronym{dcg}{DCG}{Discounted Cumulative Gain}
\newacronym{ndcg}{NDCG}{Normalised Discounted Cumulative Gain}
\newacronym{map}{MAP}{Mean Average Precision}
\newacronym{mf}{MF}{Matrix Factorisation}
\newacronym{nlp}{NLP}{Natural Language Processing}
\newacronym{cf}{CF}{Collaborative Filtering}
\newacronym{cbf}{CBF}{Content Based Filtering}
\newacronym{re}{RE}{Recommendable Entities}
\newacronym{de}{DE}{Descriptive Entities}
\newacronym{ml}{ML}{MovieLens}
\newacronym{mr}{MR}{MindReader}
\newacronym{pmi}{PMI}{Pointwise Mutual Information}
\newacronym{sppmi}{SPPMI}{Shifted Positive Pointwise Mutual Information}
\newacronym{fmf}{fMF}{Functional Matrix Factorisation}
\newacronym{als}{ALS}{Alternating Least Squares}
\newacronym{rdf}{RDF}{Resource Description Framework}
\newacronym{loo}{LOO}{Leave-One-Out}
\newacronym{rmse}{RMSE}{Root Mean Square Error}
\newacronym{knn}{kNN}{k Nearest Neighbour}
\newacronym{pc}{PC}{Pearson Correlation}
\newacronym{ac}{AC}{Adjusted Cosine}
\newacronym{hr}{HR}{Hit Ratio}
\newacronym{bpr}{BPR}{Bayesian Personalised Ranking}
\newacronym{svd}{SVD}{Singular Value Decomposition}
\newacronym{ppr}{PPR}{Personalised PageRank}
\newacronym{recsys}{RS}{Recommender System}

%% file: rules.tex

\let\oldgls\gls 
\renewcommand{\gls}[1]{\oldgls*{#1}}

\let\oldglspl\glspl 
\renewcommand{\glspl}[1]{\oldglspl*{#1}}

\newcommand{\all}{MR-ALL}
\newcommand{\bin}{MR-BIN}

\newcommand{\users}{\mathcal{U}}                 
\newcommand{\feedback}{\mathcal{C}}              

\newcommand{\entities}{\mathcal{E}}              
\newcommand{\recs}{\mathcal{E}_{rec}}               
\newcommand{\descs}{\mathcal{E}_{desc}}              
\newcommand{\labels}{\mathcal{L}}                
\newcommand{\graph}{\mathcal{G}}                 

\newcommand{\edges}{\mathcal{R}}                 
\newcommand{\relations}{\mathcal{R}}             

\newcommand{\liked}{\mathcal{E}^+}                 
\newcommand{\disliked}{\mathcal{E}^-}              
\newcommand{\unknown}{\mathcal{E}^?}
\newcommand{\mrobserved}{\mathcal{E}^\Omega}            
\newcommand{\mrunobserved}{\mathcal{E}_U}          

\newcommand{\adjacents}{\mathcal{A}_{\mrobserved}}             
\newcommand{\adjacentgroup}{\mathcal{E'}}     
\newcommand{\seedset}{\mathcal{S}}               






\newcommand{\user}{u}
\newcommand{\userother}{v}

\newcommand{\movie}{i}
\newcommand{\othermovie}{j}

\newcommand{\entity}{e}
\newcommand{\otherentity}{d}

\newcommand{\head}{h}
\newcommand{\relation}{r}
\newcommand{\tail}{t}

\newcommand{\rating}{o}
\newcommand{\seed}{s}

\newcommand{\length}{n}
\newcommand{\lengthother}{m}

\newcommand{\proplabel}{l}

\newcommand{\adjacencymatrix}{A}
\newcommand{\question}{q}

\newcommand{\ratingfunc}{R}
\newcommand{\globalpr}{S}
\newcommand{\loss}[1]{L_{#1}}
\newcommand{\Q}[1]{Q^{#1}}

\newcommand{\cvector}[1]{\mathbf{#1}}
\newcommand{\cmatrix}[1]{\mathbf{#1}}
\newcommand{\preference}[1]{\leqslant_{#1}}
\newcommand{\predpref}[1]{\:\widehat{\preference{#1}}\:}
\newcommand{\topn}{top-$n$}
\newcommand{\pipe}{\bigm|}

\newcommand{\recsys}{S}
\newcommand{\interview}{I}


 
 


%% file: commands.tex
\newcommand{\todo}[1]{\textcolor{blue}{(#1)}}

%% file: sections/abstract.tex
\glspl{kg} have been integrated in several models of recommendation to augment the informational value of an item by means of its related entities in the graph.
Yet, existing datasets only provide explicit ratings on items and no information is provided about user opinions of other (non-recommendable) entities.
To overcome this limitation, we introduce a new dataset, called the MindReader, providing explicit user ratings both for items and for \gls{kg} entities.
In this first version, the MindReader dataset provides more than 102 thousands explicit ratings collected from 1,174 real users on both items and entities from a \gls{kg} in the movie domain.
This dataset has been collected through an online interview application that we also release open source.
As a demonstration of the importance of this new dataset, we present a comparative study of the effect of the inclusion of  ratings on non-item \gls{kg} entities in a variety of state-of-the-art recommendation models.
In particular, we show that most models, whether designed specifically for graph data or not, see improvements in recommendation quality when trained on explicit non-item ratings.
Moreover, for some models, we show that non-item ratings can effectively replace item ratings without loss of recommendation quality.
This finding, thanks also to an observed greater familiarity of users towards common \gls{kg} entities than towards long-tail items, motivates the use of \gls{kg} entities for both warm and cold-start recommendations.

%% file: sections/introduction.tex

\section{Introduction}\label{sec:introduction}

The goal of \glspl{recsys} is to recommend items (e.g., products) to users based on some understanding of both the item attributes and the user preferences for them~\cite{RecommenderSystemsInCommercialUse}.
They have been widely adopted in many online systems~\cite{RecommenderSystemsInCommercialUse,NetflixDataset}.
In order to infer the preferences of a user, \glspl{recsys} require at least some information about past user-items interactions, which is usually referred to as user feedback (explicit or implicit).
Explicit feedback comes from user-item interactions where the user's intention is to provide feedback (e.g., an explicit rating)~\cite{MovieLens, NetflixDataset, ItemCoocurrenceJointLearning, SimonFunkSVD, CollaborativeDeepLearningWang15}.
Implicit feedback is collected when it is not the user's intention to actually provide any explicit judgement (e.g., a product  purchase)~\cite{hu2008collaborative}.
%
%
%
%
To enrich the available information, e.g., with additional item features, the system can  also employ a \gls{kg}~\cite{KnowledgeAwareAutoencoders, UnifyingKGsAndRecommendation}.
A \gls{kg} is a heterogeneous graph representing entities like products, people, places, and concepts as nodes, and the relationships among them as typed edges.
Hence, this unifying model combines both recommendable items (the products) and all other non-recommendable items (called descriptive entities).
It has been shown that it is possible to leverage explicit feedback from the user for some non-item objects~\cite{das2013learning}.
In particular, prior works~\cite{tagommenders,Gedikli2013} have sought to investigate the effects of incorporating ratings on item tags 
finding that both inferred and explicit ratings on tags lead to higher recommendation quality.
However, tags
\begin{enumerate*}[label=(\roman*)]
    \item have no semantic inter-relations,
    \item have no explicit semantic relationship to items, and
    \item vary greatly in quality~\cite{Gedikli2013}
\end{enumerate*}.

\emph{Instead, in this work, we address user ratings on descriptive entities from heterogeneous \glspl{kg}, overcoming the aforementioned issues (Section~\ref{sec:motivation})}.
To the best of our knowledge, no previous study has investigated the effect of including user explicit feedback over non-recommendable entities in a KG (Section~\ref{sec:related_work}).
Moreover, no public dataset exists with this kind of information.
Therefore, the first contribution of this work is the MindReader dataset (publicly available at \url{https://mindreader.tech/dataset}).
This dataset has been collected through a data collection platform that asks real users to provide ratings on both recommendable and descriptive entities in the movie domain (Section~\ref{sec:methodology}).
Hence, as a second contribution, we also release an extensible open-source platform for collecting  user rating for KG-enhanced datasets.
The analysis of this data (Section~\ref{sec:analysis}) provides a number of insights w.r.t. how user preferences correlates to various types of \gls{kg} entities.
Finally, to demonstrate whether user ratings on descriptive entities can be beneficial in generating more personalised recommendations,  we investigate the effect of their inclusion in a large set of state-of-the-art machine learning models for personalised recommendation (Section~\ref{sec:expevaluation}). 
Thanks to this new dataset, as a third contribution, we provide some initial findings that explain how the inclusion of ratings on descriptive entities affects the quality of recommendations.
Among others, our results suggest that descriptive entity ratings can serve as replacements for recommendable entity ratings in the warm-start setting, and motivate their utility also in the cold-start setting.



%% file: sections/motivation.tex

\section{KG-guided recommendations}\label{sec:motivation}
Consider the case of movie recommendations.
When modelling user preferences, we are concerned with what type of content users prefer.
For instance, whether they like Science Fiction, dislike Horror, or prefer movies by some specific director.
Some of these preferences can be combined in different ways, e.g., a user could generally like biographies but generally dislike science fiction, as shown in \autoref{fig:motivation}.
Yet, in inferring such complex preferences, observations on recommendable entities alone can be insufficient.
Additionally, to infer user preferences from a small set of user-entity observations, eliciting feedback towards other, non-recommendable, entities (e.g., actors and genres instead of movies) can intuitively be more informative.
Finally, the relationships between an actor and a movie and between a genre and a movie are drastically different, hence user preferences towards an actor or a genre have different effect on the user preference towards the movie.
For example, if we know only that the user dislikes the movie "Cloud Atlas" starring Tom Hanks, inferring that the user dislikes both science fiction and movies starring Tom Hanks, would prevent us from recommending "Catch Me If You Can" although the user likes biographies.
Without an appropriate modelling we cannot imagine to infer all these nuances.

Moreover, having access to explicit user feedback on descriptive entities is particularly important for the cold-start settings~\cite{CoachedConversationPreferenceElicitation, TowardsConversationRecommenderSystems}. 
In a cold-start scenario, we are provided with a set of user preferences for items and a new unseen item is added to the database.
In this case, it is possible to transfer the information from descriptive entities to the new unseen item.
The complementary cold-start setting is a new user accessing the system.
In this setting, a common strategy is instead to conduct an interview with the user to determine their preferences~\cite{CoachedConversationPreferenceElicitation, TowardsConversationRecommenderSystems}.
If one were to most quickly determine a user's movie preferences from a blank slate, it makes little sense to immediately ask towards a specific set of movies~\cite{CoachedConversationPreferenceElicitation, TowardsConversationRecommenderSystems}.
Among others, there is the possibility that the user might not be familiar with the movies they are asked about.
Instead, since users generally have an opinion about Horror movies, even when not familiar with many movies of that genre, asking about their preference towards the genre is more likely to provide reliable information.
Therefore, in this work we provide the first dataset with user-entity ratings both for recommendable and descriptive entities.
This will allow the study of methodologies to develop novel KG-guided recommendation systems to address all the issues mentioned above.

%% file: sections/related.tex

\section{Related work}\label{sec:related_work}
\input{Tables/related.tex}
A multitude of datasets for personalised recommendation exists~\cite{MovieLens, NetflixDataset, BookCrossing, LastFM}, and several works have considered using \glspl{kg} for preference elicitation~\cite{KnowledgeAwareAutoencoders, UnifyingKGsAndRecommendation}.
Common for all these existing datasets is that they only contain explicit user ratings on recommendable entities.
Few works have instead explored the advantages of eliciting explicit user feedback on objects constituting non-recommendable items, in particular product tags~\cite{tagommenders,Gedikli2013} and other item features~\cite{das2013learning,CoachedConversationPreferenceElicitation,MouraoRochaKonstanMeira2013}.
Yet, none of the existing datasets provide ratings for KG entities that are non-recommendable items.

\para{Ratings for tags.}
\emph{Tagommenders}~\cite{tagommenders} provides ratings by $995$ MovieLens users for $118,017$ movie tag.
Here, tags comprise a set of labels that are generally ascribed to genres and categories (e.g., ``french movies'', ``terrorism'', ``love story'').
Hence, the effects of tag-based \glspl{recsys} are evaluated in comparison to recommendable-entity-based \glspl{recsys}, finding that tag-based \glspl{recsys} using explicit tag ratings outperform recommendable-entity-based \glspl{recsys} that rely on \gls{cf} methods.
Furthermore, a linear combination of the best tag-based \gls{recsys} and the best recommendable-entity-based \gls{recsys} is demonstrated to yield the best results.
%
Another work proposes \emph{regress-tag}~\cite{Gedikli2013} collecting explicit user ratings for specific movie-tag pairs over $\sim$\,$100$ movies and $19$ users, for $5,648$ ratings in total.
Thus, this dataset collects multiple ratings from the same user for the same tag, depending on the movie to which it is assigned.
The results show that using explicitly collected tag ratings provides better performance than implicitly derived tag ratings.
Finally, \emph{ShoppingAdvisor}~\cite{das2013learning} takes a different approach to preferential data, and collects two datasets of explicit and implicit user feedback on cars and cameras, respectively.
In practice, keywords appearing in reviews and product tags are extracted and used to describe user-product relationships.
This association is inferred as a positive relationship.
For instance, a camera can be tagged as ``food'' and ``nature'' if a given user has used those tags on an picture taken with the camera, and should be recommended to a user looking for a camera to take pictures of food or nature.
As for the ground-truth ranking data, popularity is used as a proxy for ground-truth ratings, ranking the cameras by the number of pictures taken with a given camera for every single tag.
Although the ShoppingAdvisor datasets contain a large variety of such tags on the recommendable entities~\cite{das2013learning}, they suffer from the same limitations as Tagommenders~\cite{tagommenders} as the tags are free-text and added by users without restriction.

\emph{All the above works differ to MindReader because the ratings are collected on user-provided tags rather than on entities in a reference knowledge graph}.
Furthermore, none of these work addresses the effect of substituting movie ratings with tag ratings, leaving the informational value of tag-ratings - whether explicit or inferred - unclear.
Conversely, we aim at determining the informational value of explicit descriptive entity ratings, and we conduct experiments to address this issue directly.
Furthermore, although the tags are plentiful and can potentially serve as useful metadata in a recommendation setting, there is only one type of relationship between a tag and a recommendable entity.
Hence, also in this case, the tags are not inter-related as entities may be in a \gls{kg}.
Finally, none of these datasets are publicly available.
\emph{MindReader alleviates these issues by collecting ratings on entities} - recommendable as well as descriptive - drawn from an existing \gls{kg} where the inter-relations between entities and the semantics of those relations are explicitly defined.
Moreover, by representing entities in a \gls{kg} we can exploit also (direct and indirect) relationships connecting descriptive entities to recommendable items, a type of information that is completely missing for tags.

\para{Ratings for item features.}
Recent work on preference elicitation processes~\cite{CoachedConversationPreferenceElicitation} highlighted the importance of eliciting user preferences on \textit{rich properties} of items rather than only the items themselves, proposing a "Wizard-of-Oz"-like methodology for effective preference elicitation through dialogue.
The user is asked first \textit{what sort of movies} they like, then for an \textit{example} of such movies, and finally \textit{what in particular} was appealing about the provided example movie.
Such rich properties of items can be considered analogous to descriptive entities.
Their finding supports the potential value of explicit observations from users on descriptive attributes in personalised recommendation tasks and therefore motivates our study of explicit rating on descriptive entities.
%
A similar approach compares metadata-based methods to rating-based methods~\cite{RecommendingNewMovies}. 
Contrasting to the study on preference elicitation~\cite{CoachedConversationPreferenceElicitation}, 
this work finds that even a few ratings provide better predictive power than purely metadata-based methods, surprisingly indicating that descriptive feature are less informative than the implicit evidence inferred from movie ratings.
Yet, the movie metadata adopted in this work is based purely on word co-occurrence in the movie descriptions.
Moreover, user preferences towards these attributes are synthetically inferred from movie ratings as no explicit feedback on words is provided.
This differs from our setting where we instead collect explicit feedback on both items and descriptive entities connected to them (directly and indirectly)
%
Finally, a different approach investigates how \emph{non-content attribute preferences} affect hybrid \glspl{recsys}~\cite{MouraoRochaKonstanMeira2013}.
In this case, non-content attributes, such as popularity, recency, and similarity to a user previous interactions are leveraged to explain the reason behind some user item selection.
Since non-content attributes are derived from item metadata, user sensitivity towards a specific attribute is inferred and not explicitly provided.
Interestingly, \gls{cf} methods are not able to derive preferences from non-content attributes, while hybrid models can exploit these preferences for increased performance.
In our work, we also consider how hybrid models can make use of both the entity ratings and the \gls{kg} entities in generating useful recommendations.

\para{Other data collection platforms.}
Our data collection application is inspired by \emph{Tinderbook}~\cite{tinderbook}, which provides book recommendations based on few binary ratings provided by the user.
Similar to MindReader, Tinderbook 
\begin{enumerate*}[label=(\roman*)]
    \item exploits an extension of state-of-the-art \gls{kg} embedding methods, and
    \item relies on an existing knowledge base (DBpedia) to obtain book information
\end{enumerate*}.
Nonetheless, while \emph{Tinderbook} constitutes a proof-of-concept of how \glspl{kg} and machine learning can be used for recommendations, no dataset is provided to validate such approaches.
MindReader instead provides a reference dataset to support the study of new methods in product recommendation based on ratings of both recommendable and descriptive entities.
A widely used dataset is MovieLens~\cite{MovieLens}, which is based on a \gls{recsys} created by the GroupLens Research group.
Like MindReader, MovieLens is a \gls{recsys} platform that is also exploited for data collection and has collected a dataset of 25 million movie ratings since 1995.
Yet, MovieLens contains ratings only on recommendable entities (movies).
Moreover, while variants of the MovieLens datasets contains tags similar to descriptive entities, these are free-text, not rated by users, and missing semantic relationships.

A comparison between existing datasets and MindReader can be found in \autoref{table:related}.
In summary, prior works have focused primarily on explicit ratings on user-provided tags, which differs from the descriptive entities we consider in this work as these are provided by an existing and structured knowledge graph.
Moreover, existing works try to derive ratings to descriptive entity ratings from recommendable entities.
Yet, none of the existing datasets provide explicit ratings on non-recommendable items.
A gap that we overcome with the MindReader dataset.
Different from existing works, we build the data collection platform MindReader and collect a dataset explicit descriptive entity ratings and study their integration for recommending recommendable entities. 
Moreover, we not only support explicit \emph{like} and \emph{dislike} ratings, but we also collect explicit \emph{unknown} ratings for those cases in which users were not able to provide a rating (\emph{explicit unknowns}, e.g., a movie not seen, an unknown actor, a category of unclear semantics).
Our dataset is freely available and contains more than 100,000 ratings from 1,174 users over 10,000 entities connected with rich semantic relationships.


%% file: Tables/related.tex
\newcommand{\cding}[2]{\textcolor{#1}{\ding{#2}}}
\newcommand{\yes}{\cding{myGreen}{52}}
\newcommand{\no}{\cding{red}{55}}

\begin{table*}[]
\begin{adjustbox}{max width=\linewidth}
\begin{tabular}{lrrrrllcccc}

\textbf{Dataset} & \textbf{\#Users}   & \textbf{\#Entities} & \textbf{\#Relation types} & \textbf{\#Ratings} & \textbf{Domain} & \textbf{Feedback} & \textbf{Explicit} & \textbf{Non-items} & \textbf{Explicit unknowns} & \textbf{Available} \\ 
\hline
\hline
Gedikli et. al~\cite{Gedikli2013}   & 19                 & 100                 & 1                          & 5,648              & Movies          & 5-star            & \yes              & \yes                & \no   & \no \\ 
MovieLens-100K~\cite{MovieLens}   & 610                & 9,724               & 0                          & 100,836            & Movies          & 5-star            & \yes              & \no                 & \no   & \yes \\
Tagommender~\cite{tagommenders}      & 995                & 9,724               & 1                          & 118,017            & Movies          & 5-star            & \yes              & \yes                & \no   & \no \\
CCPE-M~\cite{CoachedConversationPreferenceElicitation}          & 502            & 4,259             & 3                          & 6,297         & Movies           & Conversational        & \yes               & \yes                 & \no   & \yes \\
Netflix~\cite{NetflixDataset}          & 6,769              & 7,026               & 0                          & 116,537            & Movies          & 5-star            & \yes              & \no                 & \no   & \yes \\ \hline
\rowcolor{myBlue!10} 
\textbf{MindReader}       & \textbf{1,174}              & \textbf{10,030}              & 8                          & \textbf{102,160}            & \textbf{Movies}          & \textbf{Like/Dislike/Unk.}       & \yes              & \yes                & \yes  & \yes \\ \hline
Das et. al (Cars)~\cite{das2013learning}& 2,180              & 606                 & 1                          & 2,180              & Cars            & Binary              & \no               & \yes                & \no   & \no \\ 
Das et. al (Cameras)~\cite{das2013learning}& 5,647           & 654                 & 1                          & 11,468             & Cameras         & Binary              & \no               & \yes                & \no   & \no \\ 
LibraryThing (DBPedia)~\cite{tinderbook}          & 6,789            & 9,926             & 11                          & 410,199         & Books           & 10-star        & \yes               & \no                 & \no   & \no \\
BookCrossing~\cite{BookCrossing}     & 278,858            & 271,379             & 0                          & 1,149,780          & Books           & 10-star           & \yes              & \no                 & \no   & \yes \\ 
Last.fm~\cite{LastFM}          & 359,347            & 186,642             & 0                          & 17,559,530         & Music           & Play count        & \no               & \no                 & \no   & \yes \\

\end{tabular}
\end{adjustbox}
\caption{Characteristics of MindReader and existing datasets.}\label{table:related}
\vspace*{-12pt}
\end{table*}

%% file: sections/methodology.tex

\section{Methodology}\label{sec:methodology}




To collect the \emph{MindReader dataset}, we developed and published a web-based application called MindReader (\url{https://mindreader.tech/}) wherein we follow a gamification approach to elicit user preferences through crowdsourcing over the entities of a movie knowledge graph called  the  MindReader \gls{kg}.
In MindReader, users are asked about their preferences on movies and possibly related descriptive entities such as actors and genres.
At the end, the system tries to guess some user preferences so that users are provided with a list of recommendations as a ``reward'' for playing the game.
Here, we describe how we constructed the MindReader \gls{kg} (\autoref{sec:kg_construction}) and we outline the overall flow of the data collection application (\autoref{sec:mindreader_phases}).
This application allowed us to collect more than $100,000$ explicit ratings, which we analyse later (\autoref{sec:analysis}).

\subsection{Knowledge graph construction}\label{sec:kg_construction}

\glspl{kg} model how different entities, from one or more domains, are connected, i.e., it models heterogeneous entities and their semantic relationships (as in \autoref{fig:motivation})~\cite{IndustryKG}.
Within the movie domain, entities could be movies, genres, and actors.
A \gls{kg} is then a labelled directed multigraph represented as the triple $\langle\entities, \edges, \labels\rangle$.
Hence, in a \gls{kg} the entities are represented with nodes $\entities$, names for entities and relationships with labels $\labels$, and the relationships are from the set of edges ${\edges}{\subseteq}{\entities}{\times}{\labels'}{\times}{\entities}$, where ${\labels'}{\subseteq}{\labels}$ captures the relationships types, such that there exists a mapping ${\Phi}:{\entities}{\cup}{\edges}{\mapsto}{\labels}$.
Moreover, we say that the set of entities $\entities$ is composed by two sets: one of \emph{recommendable entities} $\recs$ (e.g., movies) and the other of \emph{descriptive entities} $\descs$ (e.g., actors, or genres), such that ${\entities}={\recs}{\cup}{\descs}$.

We constructed the MindReader \gls{kg} over a subset of $9,000$ movies from the MovieLens-100K~\cite{MovieLens} dataset (as of April 2020).
Although larger versions of the dataset are available (e.g.,  MovieLens-25M~\cite{MovieLens} with 62k movies), we started with a smaller version, which is still among the most used in the literature~\cite{jhamb2018attentive, zhang2017autosvd++} to ensure higher coverage of entities (and we plan to employ larger versions in our next version of the dataset). 
Moreover, even though the MovieLens dataset contains explicit user ratings, which are useful for traditional recommenders and popularity-based sampling, it is otherwise limited in terms of additional metadata to use as descriptive entities.
Specifically, it contains only the release year for each movie as well as a number of user-generated tags.

To extend the \gls{kg} with descriptive entities, we linked the movies in the MovieLens dataset with those in Wikidata~\cite{wikidata} by means of existing shared identifiers (e.g., IMDb IDs, \url{https://www.imdb.com/}).
Then, for every movie, we obtained from Wikidata its associated actors, directors, production studios, release decade, genres, and main subjects.
As the hierarchy of genres and subjects is included in Wikidata, we include that information in the \gls{kg} as well.
Finally, to maintain a higher graph density, we deleted all entities with only one relationship.
A sample of the \gls{kg} which exemplifies the inclusion of hierarchical information can be seen in \autoref{fig:motivation}. 
The final \gls{kg} is stored in the Neo4j (\url{https://neo4j.com}) graph database.
We summarise the graph statistics in \autoref{tab:graph_stats}.

\input{Figs/first_triple}

\subsection{The phases of MindReader}\label{sec:mindreader_phases}
To obtain ratings for various types of entities, we design the MindReader application as a game where the user is asked to express an explicit preference toward various entities in the underlying \gls{kg} with the final goal of receiving a list of movie recommendations.
Entities are presented in batches, so that the user is asked to rate an entire batch before moving to a new batch.
Moreover, we want to obtain at least some ratings about entities (both recommendable and descriptive) that are connected one to the other.
Therefore, the game contains three phases (see \autoref{fig:flow}):
\begin{enumerate}[leftmargin=*]
    \item \label{initial} An initial phase where the user is asked exclusively about recommendable entities.
    \item \label{exploration} An exploration phase where the user is iteratively asked about entities (both recommendable and non-recommendable) potentially related to those they have stated preference on. 
    \item \label{final} A final recommendation phase where system tries to guess the user preferences.
\end{enumerate}

In transitioning between these phases, the application maintains a state for each session and each user.
In all phases, whenever an entity is presented to the user, the user can provide three explicit feedbacks: \emph{like}, \emph{dislike}, or \emph{unknown}.
In the state representation, the system keeps track the entities the user likes ( $\liked$), dislikes ($\disliked$), as well as those entities the user states they do not know how to rate ($\unknown$), with $\mrobserved =\liked \cup \disliked \cup \unknown$ being all entities rated by the user.
Also, we ensure the user is asked to rate an entity at most once.


During phases \ref{initial} and \ref{exploration}, the user is shown one or more batches of $9$ entities.
The small number has been selected to keep the list of entities to rate short and the feedback process enjoyable for the user.
After each batch of entities is rated completely, the session and user states are updated then, if the phase termination criteria is met (see \autoref{fig:flow}), the application moves to the next phase, otherwise the same phase continues with a new batch.

\para{Initial phase.}
The purpose of the initial phase is to gather feedback on a seed set of recommendable entities, i.e., only recommendable entities are shown.
The termination criteria for the initial phase is $\liked\cup\disliked\neq\emptyset$, after which the interview proceeds to the exploration phase (phase~\ref{exploration}).
When sampling the entities in the initial phase we attempted to optimise the probability that the user knows at least one of the entities in the list.
As a proxy for popularity, we used the number of ratings provided in the MovieLens dataset.
The rating distribution of MovieLens is skewed towards movies from the mid-1990s, which some preliminary analysis showed to be caused by a large number of MovieLens users not being active in the platform in recent years.
%
%
%
%
%
While older movies with many ratings are indeed popular, users usually have a fresher memory regarding more recent movies. 
As such, we conduct movie sampling while both considering popularity and recency.
Thus, for a movie $\movie \in \recs$, we define a weight function
$W(\movie) = |Ratings(\movie)| * \max(1, Age(\movie) - 2000)$
where $Ratings(\movie)$ are the ratings for $\movie$ in the MovieLens dataset, and $Age(\movie)$ is the number of years since the release of $\movie$.
When sampling a movie $\movie$ for user $\user$, this is sampled with probability proportional to 
$W(i)$ normalised over the sum of weights of all the movies not already rated by the user.
While different sampling criteria could be adopted, our initial analysis with few test users suggested this formula to provide a reasonable balance between popularity and recency so to reduce the number of unknown movies.

\para{Exploration phase.}\label{exp_phase}
In the exploration phase, we collect explicit ratings both for non-recommendable and recommendable entities.
%
The termination criteria for this phase is that the user must have liked or disliked at least $30$ entities of any type, i.e., $|\liked \cup \disliked|\geq 30$, after which the interview proceeds to the recommendation phase. 
%
The connections in the MindReader \gls{kg} are used to determine which entities the user should be asked about in this phase.
These are entities that are adjacent (i.e., at 1-hop distance in either direction) to those in $\disliked$ and $\liked$.
In order to diversify the entities asked about, each round has three sets of $3$ entities sampled independently respectively among those adjacent to $\disliked$, to $\liked$, and to a different set of random recommendable entities sampled again based on the weighting of $W$. 
%
Among all the entities adjacent to the given seed set,  we conduct a weighted sampling based on their global PageRank~\cite{OriginalPageRank} in order to elicit ratings for more popular entities.
Moreover, to balance coverage of entities of all types, we split the adjacent entities into their respective types (e.g., genre, decade, actor) and randomly sample from each of these splits.

\vspace*{-4pt}
\para{Recommendation phase.}
In the recommendation phase we generate two lists of recommendable entities for the user based on their previous answers: one with entities that we guess they like, and one with entities that we guess they dislike.
We then ask the user feedback on all these recommendable entities.
The main purpose of this phase is not to provide recommendations of high quality, but rather to collect more ratings.
Furthermore, by providing recommendations to the user as a form of reward, we hope to motivate continued use of the application.  
To determine which entities to recommend, we follow a sampling approach similar to the one used in the exploration phase.
Given the entities $\liked$, we consider their adjacent recommendable entities and take the top-$25$ entities by number of connections to entities in $\liked$.
This number has been selected in order to reduce computational load.
From this subset, we sample recommendable entities by weight-based sampling on global PageRank as described previously.
We repeat then the same process for entities in $\disliked$.
We opted for this computationally simple approach since producing high quality recommendations is not the application primary goal.
Moreover, to increase coverage and avoid recommendations only related to the initial seed, an additional set of pseudo-randomly sampled entities is presented to the user.
When all rating are collected, we thank the user for their participation and invite them to take the interview again. 
All subsequent interview answers will be associated with the same user based on a token stored in their browser's local storage.
This allows us to avoid storing any user personal information except for their preferences.

%% file: Figs/first_triple.tex
     \begin{figure}
        \includegraphics[width=\columnwidth]{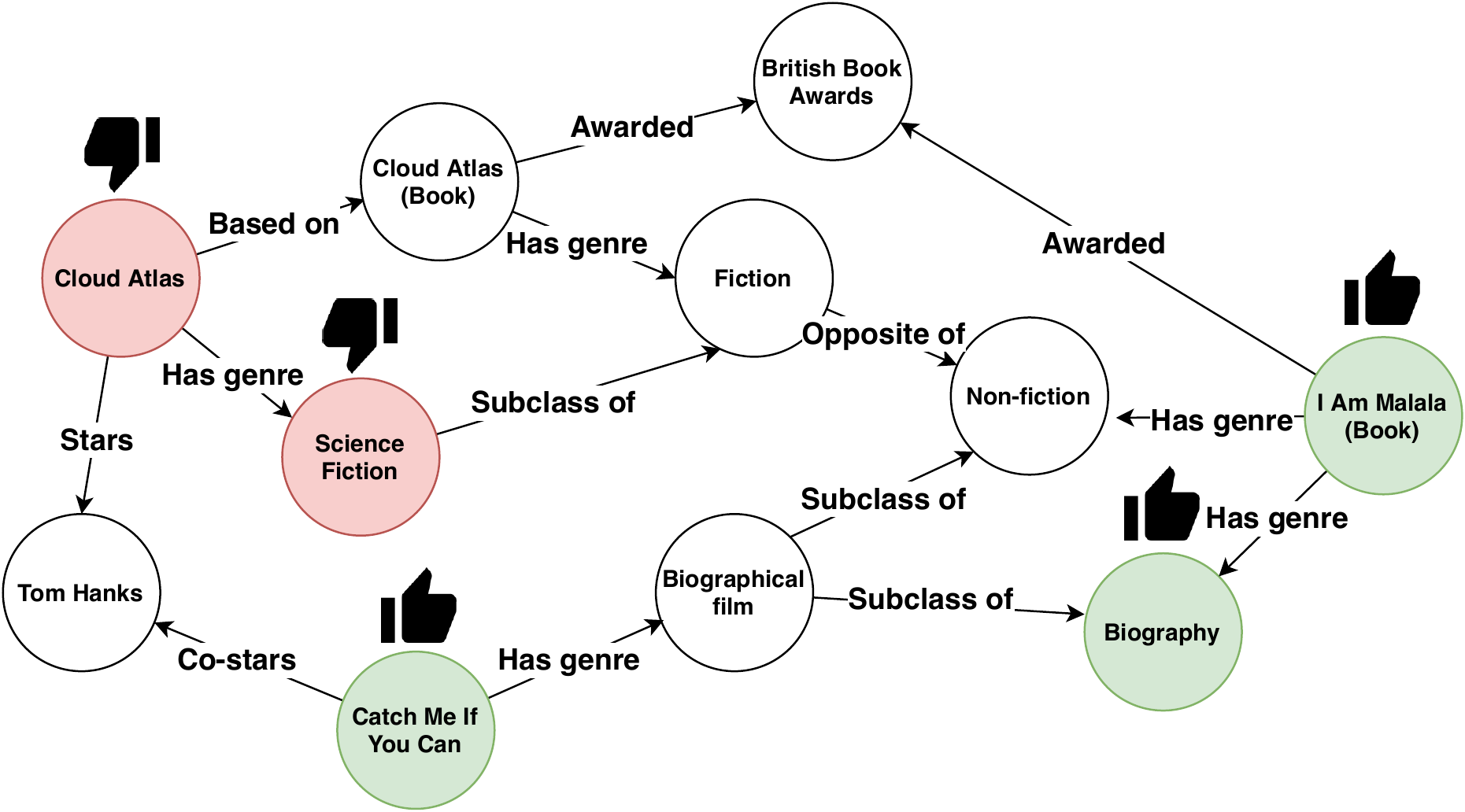}
        \Description{A graph-representation of user-entity observations. The user likes Science Fiction, Action, Interstellar and Star Wars, is partial to the Actor Samuel L. Jackson, and dislikes Quentin Tarantino, Pulp Fiction, and Django Unchained.}
        \caption{Sample graph with ratings from an hypothetical user.}
        \label{fig:motivation}
   \end{figure}
    \begin{figure}
        \centering
        \includegraphics[width=0.8\columnwidth]{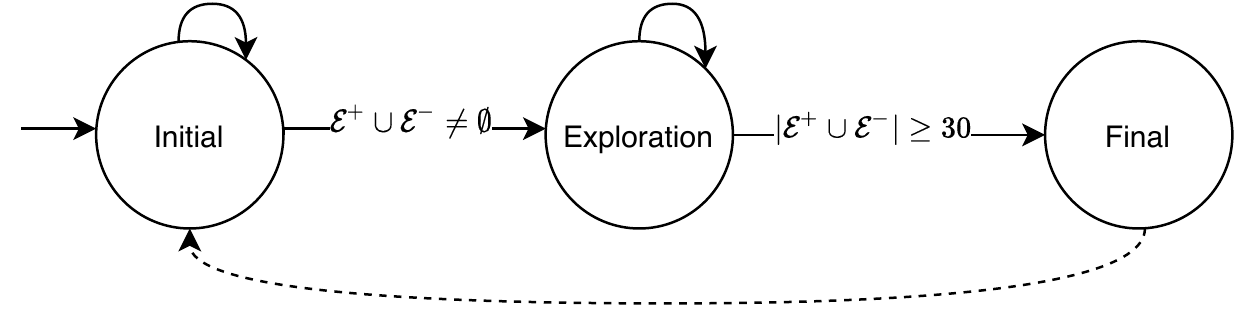}
        \Description{Flow of the MindReader application. After completing the initial phase, the user moves into the exploration phase, where they stay until the system has received at least 25 like/dislike observations. The system then moves on to the final phase, where the user is provided with two lists of movies we expect the user to like and dislike, respectively. The user can choose to restart the game, returning to the initial phase.}
        \caption{Flow of the MindReader application. The dashed line indicates that users are able to restart the interview upon completion.}
        \label{fig:flow}
        \vspace*{-12pt}
        \end{figure}


%% file: sections/analysis.tex

\section{Analysis}\label{sec:analysis}

In this section we analyse the data collected from the MindReader application.
We present both the characteristics of the graph and the ratings we obtained.
We highlight that this is the first version of the dataset and our plan is to continue with data collection so to release larger datasets over time.
We are only considering ratings from users who completed the quiz till the final phase.
We consider two variants of our dataset: \all{} which contains all observed ratings, and \bin{} which is a subset of \all{} in which we only consider the binary like/dislike ratings.

\para{Knowledge graph statistics.}
The graph contains more than 18 thousands nodes and 198 thousands edges (\autoref{tab:graph_stats}).
It includes edges from movies (the recommendable entities) to its adjacent descriptive entities as well as among descriptive entities in a hierarchy to their superclasses.
%
All nodes are directly related to at least $4$ other entities in the \gls{kg}, and we can infer that all entities are related to a fair number of neighbouring entities (median degree is $10$).
Moreover, we see that there exist a number of entities that are highly connected since the maximum degree is more $4.4$ thousands and the average degree is $21$.
Finally, there is only one connected component in the \gls{kg}.
That is, there is an undirected path from any entity to any other entity in the \gls{kg}.

\para{Long tail rating distribution.}
The MindReader dataset presents a classical long-tail distribution of ratings (\autoref{fig:distribution}), which is similar to that of the original MovieLens dataset.
Therefore, a small fraction of entities are popular -- and receive most positive ratings (the short head) -- while the remaining majority received fewer ratings and fewer preferences as well (the long tail).
In \autoref{fig:distribution}, the entities on the vertical axis are sorted by popularity with the most popular entities at the bottom, i.e., at $1\%$ is the top-$1\%$ most popular entities by number of ratings.
Therefore, for the \bin{} dataset, we observe that $20\%$ of ratings involve $1.98\%$ of the most popular entities ($\approx 57$ entities).
%
Note that long-tail distributions are usually common, and for this reason trivial approaches to recommend the \topn{} entities in the short head are usually a competitive baseline~\cite{SamplingAndItemKnn}.

\para{Co-rated entities.}
\input{Figs/third_triples}
In order to test collaborative filtering (\gls{cf}) approaches, we would require that users with similar rating patterns like similar entities~\cite{MouraoRochaKonstanMeira2013}.
For testing such methods then, it is necessary to evaluate entity co-ratings.
Therefore, we consider the number of entities rated by the same user-pair, i.e., for a user-pair $\{u,v\}\in \{\{u,v\}| u,v\in \users \text{ and } u \neq v\}$ we consider the number of entities that both user $u$ and $v$ have rated.
Having many user-pairs with a high number of co-rated entities theoretically increases the performance of \gls{cf} methods.
%
Also, we provide this analysis only for \bin{}, since ``don't know'' ratings would not be commonly used in \gls{cf} approaches (see \autoref{fig:mr-re-co-ratings} and \autoref{fig:mr-de-coratings}).
%
While users of MovieLens have a minimum of $25$ ratings and a mean of $165$ ratings per user, the MindReader game stops in few steps with a mean of $31$ ratings per user ($87$ if we consider ``don't know'' ratings). 
For this reason co-ratings for recommendable entities (REs) are generally under $20$, while co-ratings for descriptive entities (DEs) are a bit higher (mostly under $50$).
%
The reason for higher co-ratings of descriptive entities is twofold.
First, there are \glspl{de} that are very central in our \gls{kg}, e.g., the \emph{Drama} and \emph{Action} genres, and those are shown to the users more frequently.
Moreover, as we will see later, users are more likely to provide ``don't know'' rating for  \glspl{re} than  \glspl{de}.
This is particularly important when, as we show later, we are able to infer user preference from  \glspl{de} instead of \glspl{re}.
This allows us to infer the same amount of information by asking users to rate a smaller amount of entities that are descriptive entities and for which we are more likely to obtain relevant feedback.

\para{Coverage.}
\autoref{fig:coverage} shows the coverage in terms of rating for different entity types.
As many person entities are both actors and directors, we have listed these as ``Person''.
Similarly, many subjects are also genres, hence we refer to the union of these as ``Category''. 
We include the fraction of entities for which there are no observations, the fraction of entities for which there are only ``Don't know'' observations, and the fraction of those for which there are binary ratings (i.e., at least one user liked or disliked the entity).
For entities with both a ``Don't know'' observation and a binary rating, we include it only in the fraction of entities with binary ratings.
As expected from the sampling approach we employ, coverage is generally higher for entity types with fewer entities.
The exception is movies which has almost full coverage when considering ``Don't know'' observations.
This is because movies can be sampled in all phases, while \glspl{de} are shown only in the second phase.
\para{Ratings distribution.}
%
We extract separately the recommendable and descriptive entity ratings and consider the distribution of ratings among types of entities.
The results of this analysis is shown in \autoref{fig:rating_distributions_over_labels}. %
%
%
On average, $61.8\%$ of movie ratings in an arbitrary MindReader session are ``Don't know'' ratings. 
Conversely, on average, only $21.3\%$ of genre ratings in an arbitrary MindReader session are ``Don't know'' ratings.
Naturally, even though users are not familiar with a large number of movies, they still largely have opinions on genres.
We see a similar trend for broader qualities of movies, e.g., subjects and categories, while questions towards actors, directors and film companies lead to useful feedback only in smaller proportion (this relates also to the current coverage in the dataset, as seen earlier in \autoref{fig:coverage}).
%
This is particularly important for systems that are allowed only to ask a small number of questions: they must commit to a trade-off between information gain and likelihood that they will receive a useful answer from the user.
For instance, in a cold-start scenario, the system usually conducts a brief interview with a newly arrived user, asking feedback towards specific entities to build a preference profile~\cite{TowardsConversationRecommenderSystems, CoachedConversationPreferenceElicitation}.
These interviews must both be short to keep the experience enjoyable for the user, and rich in information as to produce high-quality recommendations based on the inferred preferences of the user.
As such, in constructing the interview, the system should minimise the chance that a user is not aware of or has no opinion on a given entity they are asked about.

%% file: Figs/third_triples.tex
\begin{figure*}
    \centering
    \begin{minipage}{.32\textwidth}
        \centering
        \includegraphics[width=\textwidth]{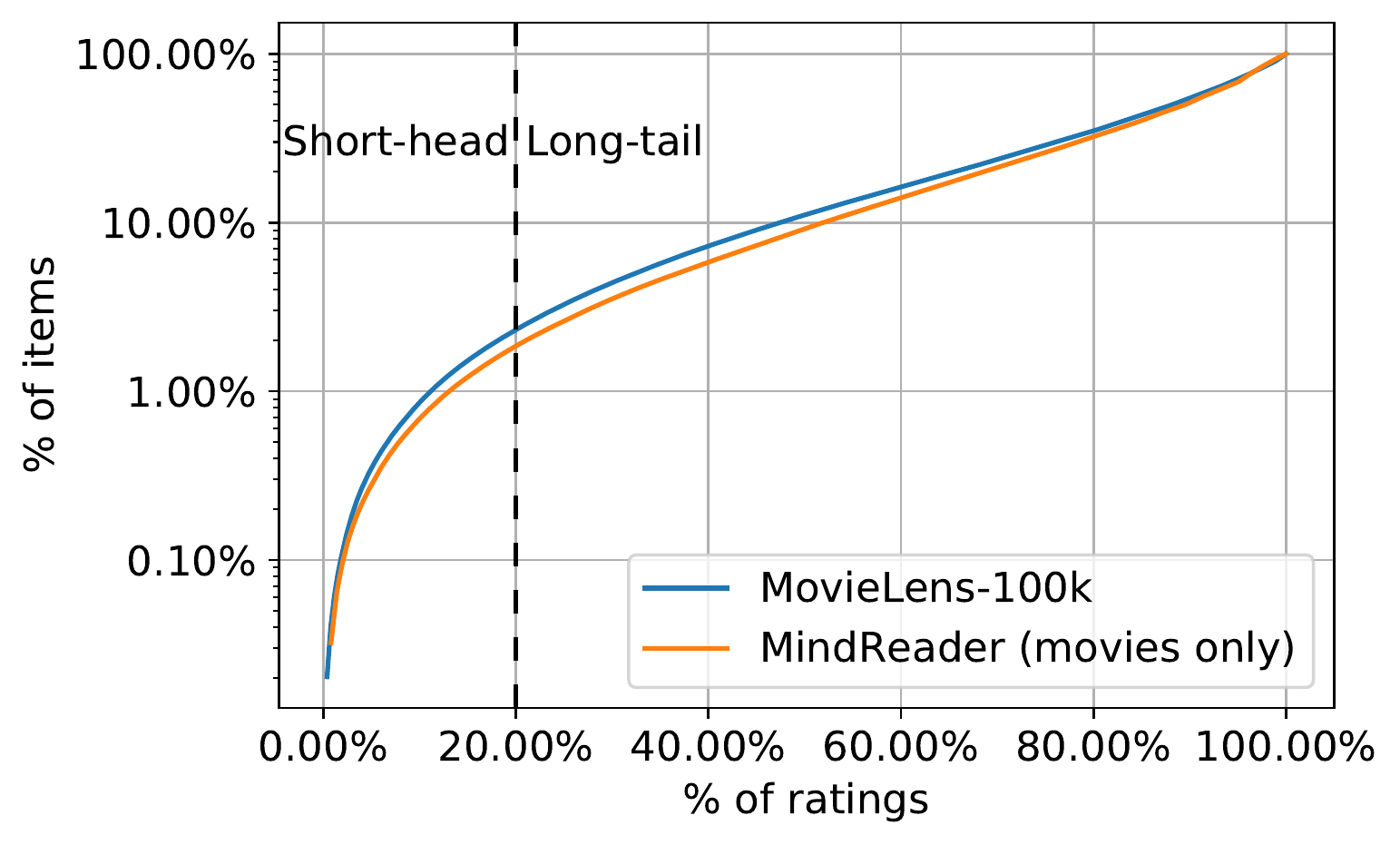}
        \caption{Rating distributions for the \bin{} and MovieLens datasets.}
        \label{fig:distribution}
        \hfill
    \end{minipage}
    \hfill
    \begin{minipage}{.3\textwidth}
        \centering
        \includegraphics[width=\textwidth]{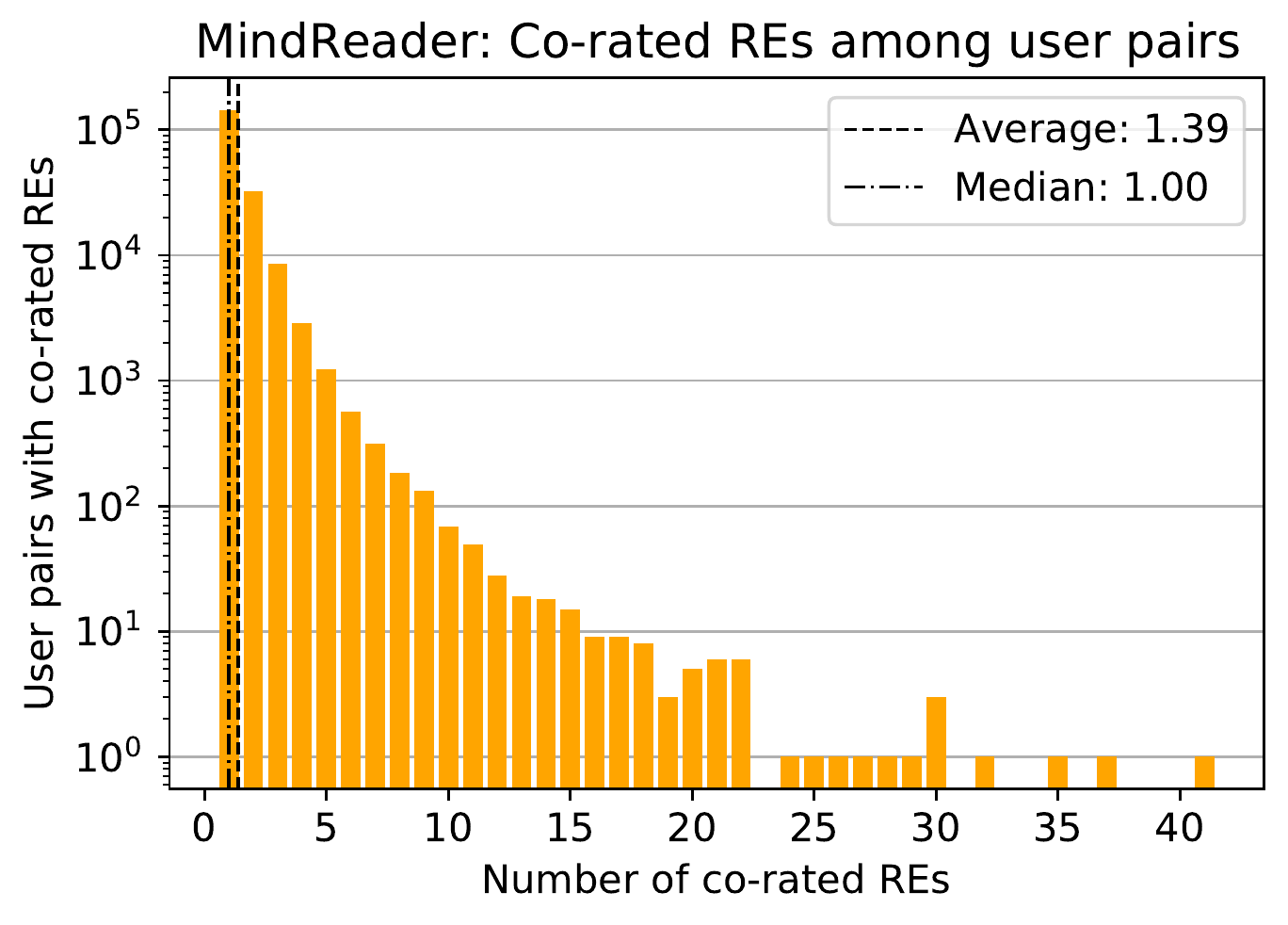}
        \vspace*{-22pt}
        \Description{MR user pairs that have co-rated a number of \glspl{re}. The x-axis is the number of co-ratings for a user pair, and the y-axis is many user pairs that share that amount of co-ratings.}
        \caption{MR user pairs that have co-rated a number of \glspl{re}.}
        \label{fig:mr-re-co-ratings}
    \end{minipage}
    \hfill
    \begin{minipage}{.3\textwidth}
        \centering
        \includegraphics[width=\textwidth]{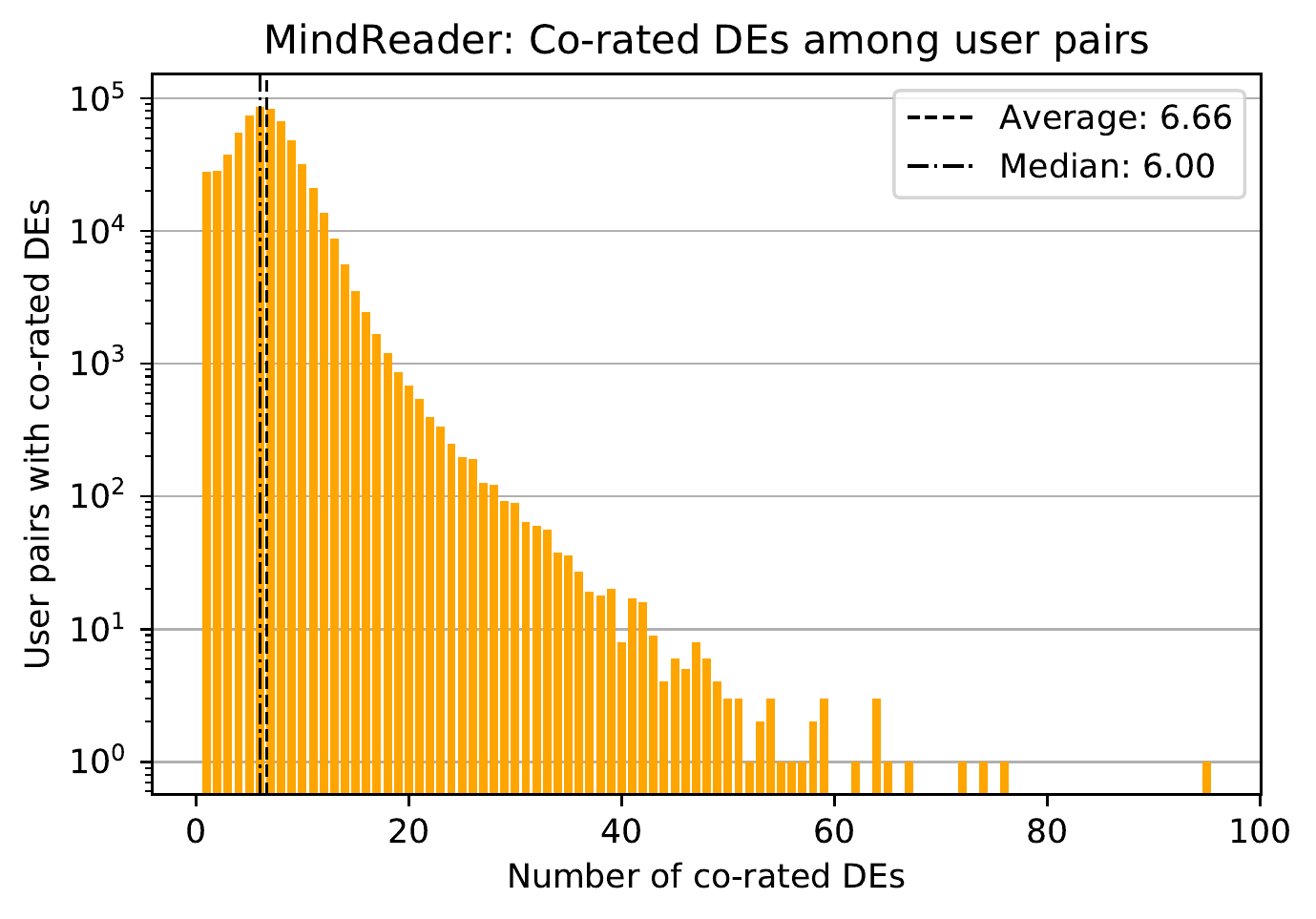}
        \vspace*{-22pt}
        \Description{A figure illustrating how many \gls{mr} user pairs that have co-rated a number of \glspl{de}. The x-axis is the number of co-ratings for a user pair, and the y-axis is many user pairs that share that amount of co-ratings.}
        \caption{MR user pairs that have co-rated a number of \glspl{de}.}
        \label{fig:mr-de-coratings}
    \end{minipage}
    \vspace*{-8pt}
\end{figure*}

%% file: sections/evaluation.tex
\section{Evaluating the Impact of Descriptive Entities}\label{sec:expevaluation}

\input{Figs/second_triple}

Now we present a preliminary evaluation of the effects of including explicit user feedback for descriptive entities in a recommendation model.
We evaluate a wide range of models comprising naive~\cite{SamplingAndItemKnn}, centrality-based~\cite{OriginalPageRank, GraphBasedCollaborativeRanking}, neighbourhood-based~\cite{SamplingAndItemKnn}, and various embedding-based~\cite{SimonFunkSVD,TransE,TransH} models for recommendation.
Specifically, we are interested in observing the effect with respect to the quality of recommendations for recommendable entities (movies) only when extending the model to include information about non-recommendable entities. 

\subsection{Evaluation setup}\label{sec:evaluation_setup}

We consider ratings as a categorical value $\in\{\text{Like}, \text{Dislike}\}$ and use data from the \bin{} dataset.
We leave the evaluation of the potential value of explicit ``Don't know'' ratings as future work.
Moreover, since this first edition of the MindReader dataset is still quite sparse in ratings, we follow the evaluation strategy of other recommender systems in presence of sparse ratings and evaluate all models using \gls{loo} evaluation~\cite{SamplingAndItemKnn,karypis2004,he2017neural}.
%
To evaluate the quality of the recommendations made, we use $k$-bounded ranking-based performance metrics~\cite{TriRankHitRateNDCGDefinitions, he2017neural}.
For every user, we generate a ranked list of the $k$ recommendable entities we expect the user to prefer the most.
%
As metric we employ \gls{hr} and \gls{ndcg} at k~\cite{TriRankHitRateNDCGDefinitions, he2017neural} and we compute them as follows:
\vspace*{-4pt}
\begin{equation}\label{eq:hr}
    \text{HR}@k = \dfrac{\textit{Number of hits @ k}}{\textit{Number of users}}
\end{equation}

%
\vspace*{-4pt}
\begin{equation}
    \text{NDCG}@k = \sum\limits_{i=1}^{k} \dfrac{2^{rel(i)} - 1}{\log_{2}(i + 1)}
\end{equation}

\vspace*{-4pt}
{\noindent}where $rel(i) = 1$ when the entity at rank $i$ is the entity the user liked and $rel(i) = 0$ otherwise.
In our evaluation, the \gls{ndcg} is normalised by definition as we only measure the metric for a single entity.
%
Therefore, higher \gls{hr}$@k$ and \gls{ndcg}$@k$ scores correlate with a higher quality of recommendations.

%% file: Figs/second_triple.tex
\begin{figure*}[ht]
    \centering
    \begin{minipage}{.3\textwidth}
        \centering
        \includegraphics[width=\columnwidth]{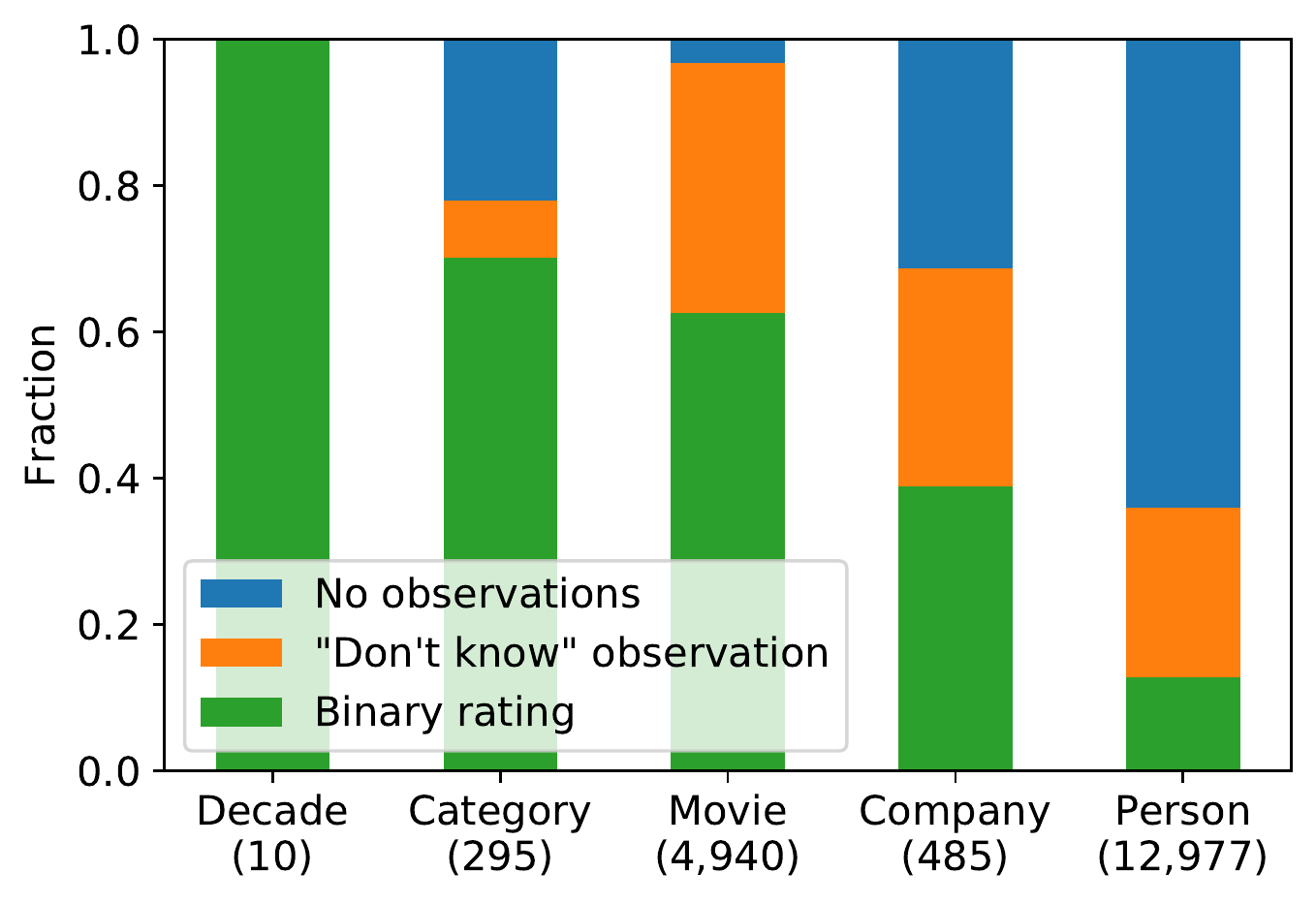}
        \caption{Coverage of different entity types from the \gls{kg}. The number of entities of the different types is shown in parentheses.}
        \label{fig:coverage}
        \hfill
    \end{minipage}
    \hfill
    \begin{minipage}{.3\textwidth}
        \centering
        \includegraphics[width=\columnwidth]{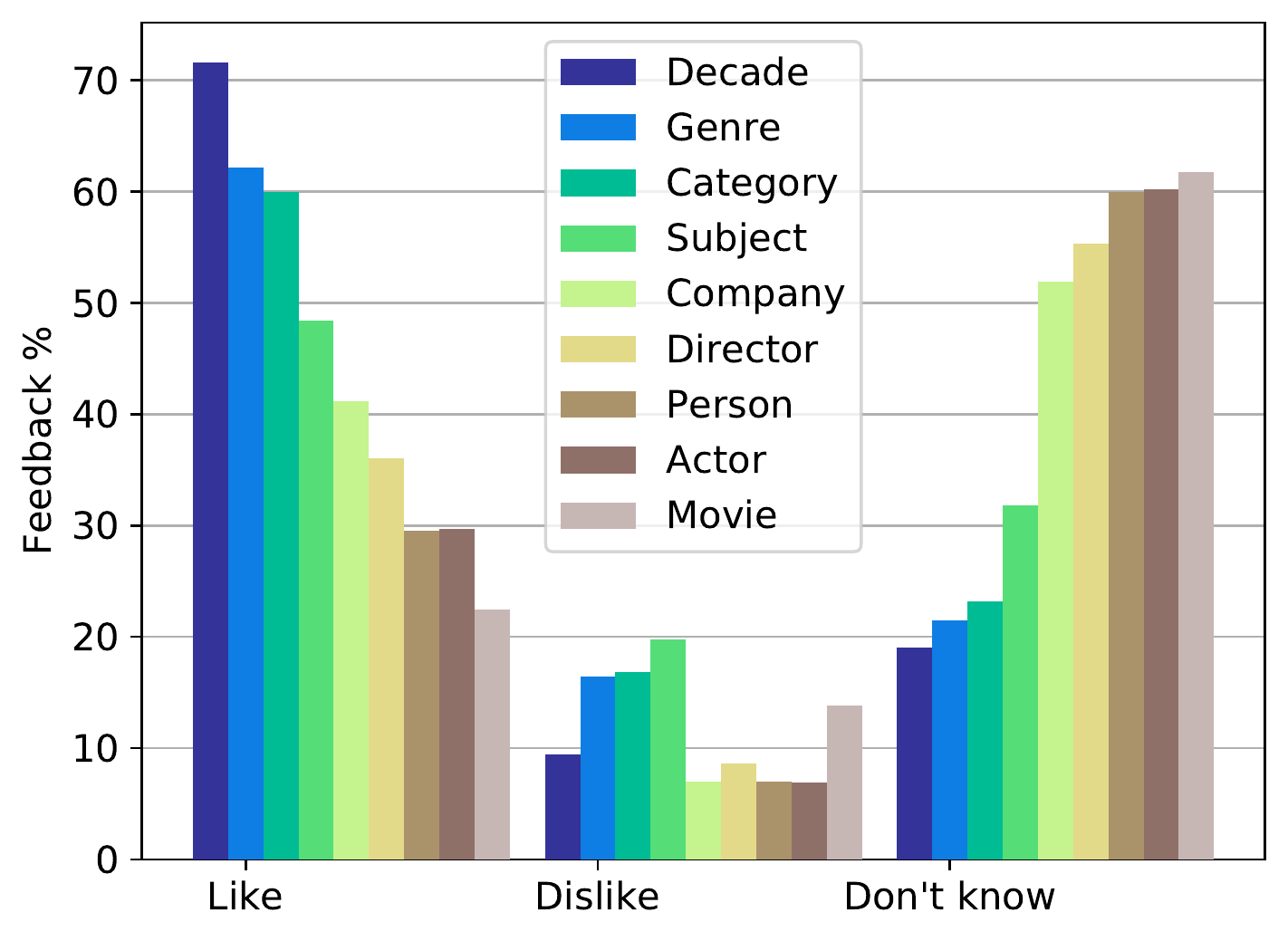}
        \vspace*{-22pt}
        \Description{A bar-chart over how ratings for all entity types are distributed over the rating categories. The feedback percentage for an entity type in a feedback category is the mean distribution over all user sessions.}
        \caption{Rating distribution over the entity categories. The feedback percentage is the mean distribution over all user sessions.}
        \label{fig:rating_distributions_over_labels}
    \end{minipage}
    \hfill
    \begin{minipage}{.3\textwidth}
                
                
                
        \begin{table}[H]
            \centering
            \renewcommand{\arraystretch}{1}
            \setlength{\baselineskip}{1\baselineskip}
            \resizebox{0.70\linewidth}{!}{%
            \begin{tabular}{l|r}
                \textbf{Measure} & \textbf{Value} \\\hline
                \# nodes & 18,707 \\\hline 
                \# decades & 10 \\
                \# companies & 485 \\
                \# categories & 295 \\
                \# movies & 4,940 \\
                \# people & 12,977 \\\hline
                \# edges & 198,452 \\\hline \hline 
                
                Minimum degree & 4 \\
                Median  degree & 10 \\
                Average degree & 21 \\
                Maximum degree & 4,454 \\\hline 
                
                \# connected components & 1 \\\hline 
                
                
            \end{tabular}}

            \caption{Statistics on the constructed \gls{kg}.}
            \label{tab:graph_stats}
        \end{table}
    \end{minipage}
    \vspace*{-12pt}
\end{figure*}

%% file: sections/experiments.tex

\subsection{Experiments}\label{sec:experiments}


Given the typical long-tail distribution in most recommendation contexts,
most recent studies have focused in designing methods able to provide non-trivial recommendations~\cite{SamplingAndItemKnn}.
Thus, similar to the literature, we perform both tests with the full dataset as well as where we remove the top 2\% of popular entities from the test set.
This allows studying the degree to which descriptive entity ratings help in generating recommendations for non-trivial entities.
%
Moreover, we compare the models over 3 different setups, namely:
%

\para{(a) Adding all ratings.}
In the first setup we study the difference in performance of the models first only on recommendable entity ratings and then when training on all entity (both recommendable and descriptive) ratings.

\para{(b) Substituting recommendable entity ratings.}
We test how well descriptive entity ratings can replace those of recommendable entities.
Thus, we train the models on datasets with a varying ratio between recommendable and descriptive entity ratings.
Specifically, let $N_{\user}$ be the number of recommendable entity ratings made by a user $u$. 
Let $M_{\user}(n), D_{\user}(n)$ be functions that sample, at random, $n$ recommendable and descriptive entity ratings, respectively, from the observed ratings of $\user$. 
Given integers $m = 4$ and $n \in \{4, 3, 2, 1\}$, we train the models on datasets containing the ratings $$\bigcup\limits_{\user} M_{\user}(\dfrac{n}{m}\cdot N_{\user}) \cup D_{\user}(\dfrac{m - n}{m} \cdot N_{\user})$$ 

\para{(c) Removing recommendable entities ratings.}
Finally, we conduct a final experiment where, using the same splits as in experiment \textbf{(b)}, we simply remove movie ratings from the training set without substituting them with descriptive entity ratings.

\para{Reproducibility.}
To support the reproducibility of our results, we publish all our experimental data, setup, and code for running all experiments.
Due to space constraints, here we report only results where we remove top popular entities from the test set and only for the first two setups.
All other results are described in the extended version of this work on \url{https://mindreader.tech/dataset/}.

%% file: sections/results.tex

\subsection{Recommendation models}\label{sec:results}

We tested the classical baseline, \textbf{TopPop}, a naive approach which recommends the recommendable entity with the most ratings.
Additional, we included two \gls{mf}~\cite{SimonFunkSVD} methods. 
The first, \textbf{MF}, a standard  model trained using \gls{als} updates, and the second \textbf{BPR}, trained by maximising the \gls{bpr} objective~\cite{BPR}.
Moreover, we also tested two methods based on kNN. 
\textbf{User kNN}, a kNN-based model recommending entities to users based on preferences of $k$ most similar users, and \textbf{Item kNN}, recommending entities to a user based on the $k$ entities most similar to those rated by the user. 

In addition, we tested 4 different variations of graph embeddings.
\textbf{TransE}~\cite{TransE}, recommending entities to users based on predicted probability of the user and entity being linked in a rating graph, trained on rating triples only (edges connecting users to rated entities).
\textbf{TransE-KG}, based on TransE, trained on both rating triples and other triples from the MindReader \gls{kg}, performing link-prediction on the MindReader \gls{kg} extended with user-rating-entity triples.
Similar to TransE, we also tested two models based on a TransH~\cite{TransH}, the base \textbf{TransH} model trained on rating triples only and \textbf{TransH-KG} including also triples from the MindReader \gls{kg}.

Finally, we tested 3 recommendation techniques based on the \gls{ppr} score~\cite{jeh2003scaling}.
The first method, \textbf{PPR-KG}, recommending movies through \gls{ppr} with the highest \gls{ppr} score navigating only the MindReader \gls{kg}, using entities that a user has liked as seeds.
The second method, \textbf{PPR-COLLAB}, navigating a \emph{collaborative graph} with nodes users and entities and edges connecting only users and their rated entities, that is without any \gls{kg} edge.
The third method, \textbf{PPR-JOINT}, finally navigates both the MindReader \glspl{kg} and the co-rating edges.

\input{Tables/ntp-everything.tex}



%% file: Tables/ntp-everything.tex
\begin{table*}[ht!]
        \centering
        \begin{adjustbox}{max width=\linewidth}
        \begin{tabular}{|l|c c|c c c c c|c c|c c c c c|}
                    \hline
                    & \multicolumn{7}{|c|}{\textbf{HR@10}} & \multicolumn{7}{|c|}{\textbf{NDCG@10}}\\\hline
                    Models & All movies & All entities & & 4/4 & 3/4 & 2/4 & 1/4 & All movies & All entities & & 4/4 & 3/4 & 2/4 & 1/4
                    \\\hline
                     BPR & $0.36 \pm 0.06$ & $0.41 \pm 0.02^*$ & & $0.39 \pm 0.01$ & $0.39 \pm 0.04$ & $0.39 \pm 0.01$ & $0.37 \pm 0.02^*$ & $0.18 \pm 0.02$ & $0.19 \pm 0.01$ & & $0.19 \pm 0.01$ & $0.19 \pm 0.02$ & $0.19 \pm 0.01$ & $0.19 \pm 0.02$
                    \\ Item kNN & $0.17 \pm 0.01$ & $0.35 \pm 0.01^*$ & & $0.17 \pm 0.01$ & $0.17 \pm 0.01$ & $0.20 \pm 0.01^*$ & $0.25 \pm 0.01^*$ & $0.09 \pm 0.01$ & $0.19 \pm 0.01^*$ & & $0.09 \pm 0.01$ & $0.09 \pm 0.00$ & $0.11 \pm 0.01^*$ & $0.15 \pm 0.01^*$
                    \\ MF & $0.42 \pm 0.01$ & $0.45 \pm 0.02^*$ & & $0.42 \pm 0.01$ & $\mathbf{0.43 \pm 0.02}$ & $\mathbf{0.42 \pm 0.01}$ & $0.41 \pm 0.02$ & $\mathbf{0.20 \pm 0.01}$ & $0.22 \pm 0.01^*$ & & $\mathbf{0.21 \pm 0.01}$ & $\mathbf{0.21 \pm 0.01}$ & $0.20 \pm 0.01$ & $0.20 \pm 0.01$
                    \\ PPR-COLLAB & $0.42 \pm 0.01$ & $0.43 \pm 0.01$ & & $\mathbf{0.43 \pm 0.01}$ & $0.41 \pm 0.02$ & $0.41 \pm 0.01$ & $\mathbf{0.44 \pm 0.01}$ & $\mathbf{0.20 \pm 0.01}$ & $0.21 \pm 0.00$ & & $0.20 \pm 0.01$ & $0.19 \pm 0.01^*$ & $0.20 \pm 0.01$ & $0.21 \pm 0.01$
                    \\ PPR-JOINT & $0.38 \pm 0.01$ & $\mathbf{0.50 \pm 0.01^*}$ & & $0.37 \pm 0.01$ & $0.36 \pm 0.02$ & $0.37 \pm 0.01$ & $0.40 \pm 0.02^*$ & $0.19 \pm 0.01$ & $0.30 \pm 0.01^*$ & & $0.19 \pm 0.01$ & $\mathbf{0.21 \pm 0.01^*}$ & $\mathbf{0.23 \pm 0.01^*}$ & $\mathbf{0.26 \pm 0.01^*}$
                    \\ PPR-KG & $0.25 \pm 0.01$ & $0.46 \pm 0.01^*$ & & $0.25 \pm 0.01$ & $0.29 \pm 0.02^*$ & $0.32 \pm 0.01^*$ & $0.38 \pm 0.02^*$ & $0.15 \pm 0.01$ & $\mathbf{0.32 \pm 0.01^*}$ & & $0.15 \pm 0.01$ & $0.19 \pm 0.01^*$ & $0.21 \pm 0.01^*$ & $0.25 \pm 0.01^*$
                    \\ TopPop & $\mathbf{0.43 \pm 0.01}$ & $0.42 \pm 0.01$ & & $\mathbf{0.43 \pm 0.01}$ & $0.42 \pm 0.02$ & $0.40 \pm 0.01^*$ & $0.40 \pm 0.02^*$ & $\mathbf{0.20 \pm 0.01}$ & $0.20 \pm 0.00$ & & $0.20 \pm 0.01$ & $0.19 \pm 0.01$ & $0.19 \pm 0.01^*$ & $0.19 \pm 0.01$
                    \\ TransE & $0.33 \pm 0.03$ & $0.31 \pm 0.04$ & & $0.32 \pm 0.03$ & $0.34 \pm 0.01$ & $0.32 \pm 0.02$ & $0.32 \pm 0.01$ & $0.18 \pm 0.01$ & $0.17 \pm 0.02$ & & $0.17 \pm 0.02$ & $0.17 \pm 0.01$ & $0.16 \pm 0.01$ & $0.17 \pm 0.01$
                    \\ TransE-KG & $0.28 \pm 0.01$ & $0.18 \pm 0.01^*$ & & $0.39 \pm 0.01$ & $0.36 \pm 0.02^*$ & $0.35 \pm 0.01^*$ & $0.33 \pm 0.02^*$ & $0.15 \pm 0.01$ & $0.10 \pm 0.01^*$ & & $0.20 \pm 0.01$ & $0.18 \pm 0.01^*$ & $0.18 \pm 0.01^*$ & $0.17 \pm 0.02^*$
                    \\ TransH & $0.28 \pm 0.04$ & $0.32 \pm 0.02^*$ & & $0.23 \pm 0.01$ & $0.26 \pm 0.02^*$ & $0.31 \pm 0.04^*$ & $0.26 \pm 0.03^*$ & $0.15 \pm 0.02$ & $0.17 \pm 0.01^*$ & & $0.12 \pm 0.01$ & $0.14 \pm 0.01^*$ & $0.16 \pm 0.03^*$ & $0.14 \pm 0.01^*$
                    \\ TransH-KG & $0.30 \pm 0.04$ & $0.35 \pm 0.02^*$ & & $0.34 \pm 0.05$ & $0.31 \pm 0.03$ & $0.29 \pm 0.03^*$ & $0.28 \pm 0.03$ & $0.16 \pm 0.02$ & $0.18 \pm 0.01^*$ & & $0.18 \pm 0.02$ & $0.17 \pm 0.02$ & $0.15 \pm 0.01^*$ & $0.15 \pm 0.01^*$
                    \\ User kNN & $0.31 \pm 0.02$ & $0.40 \pm 0.01^*$ & & $0.28 \pm 0.01$ & $0.30 \pm 0.01^*$ & $0.32 \pm 0.02^*$ & $0.32 \pm 0.02^*$ & $0.17 \pm 0.01$ & $0.21 \pm 0.01^*$ & & $0.16 \pm 0.01$ & $0.17 \pm 0.01$ & $0.17 \pm 0.01^*$ & $0.18 \pm 0.01^*$
                    \\\hline
            \end{tabular}
        \end{adjustbox}
        \caption{\{HR, NDCG\}@10 performance. Statistically significant differences in mean performance between related experiment pairs are marked with a star (\textbf{*}).  }\label{tab:full_results}
        \vspace*{-22pt}
\end{table*}

%% file: sections/discussion.tex

\subsection{Results}\label{sec:discussion}
\vspace*{-8pt}
\para{Adding descriptive entity ratings.}\label{results:all_ratings}
The results of experiment \textbf{(a)} are shown in the ``All movies/entities'' columns of \autoref{tab:full_results}.
Models were trained only on movie ratings (the ``All movies'' column), and movie- as well as descriptive entity ratings (the ``All entities'' column). Statistical significance is determined from a paired sample $t$-test between the two settings.

First considering model performance when provided with descriptive entity ratings in addition to movie ratings, \emph{we observe that most learning models achieve a statistically significant improvement in \gls{hr}$@k$ and \gls{ndcg}$@k$ when provided with descriptive entity ratings}. 
While this improvement in performance may not be of much surprise as we are essentially increasing the amount of training data for each model, it still indicates that the model is able to infer useful information also from explicit ratings on descriptive entities.
Moreover, we observe that some models such as Item \gls{knn} are especially good at making use of the added ratings, seeing $105.8\%$ improvement in \gls{hr}@$k$.
Interestingly, we note that while the translational TransH models see a statistical significant increase in performance, the TransE models decrease, possibly due to the fact the that the TransH models are more expressive and better capture relations than TransE models.


Considering the \gls{hr} and \gls{ndcg} on movie ratings alone in \autoref{tab:full_results}, we observe that TopPop achieves the highest \gls{hr}.
Other works in the literature have also shown that simply recommending the most popular recommendable entities to users is a surprisingly effective strategy~\cite{SamplingAndItemKnn}. 
Nevertheless, while TopPop performs well,  when we include descriptive entities the ranking quality by the \gls{kg}-based \gls{ppr} models, among others, becomes far superior to this baseline.
%
%
In fact, PPR-JOINT and PPR-KG are the best performing models when provided with descriptive entity ratings in \gls{hr} and \gls{ndcg} respectively when compared to all other methods. 

\para{Substituting with descriptive entity ratings.}\label{results:substituting_ratings}
The results of experiment \textbf{(b)} are shown in the ``n/4'' columns of \autoref{tab:full_results}.
In each table, all listed models were trained on datasets with shifting ratios between the number of movie and descriptive entity ratings, where the ``4/4'' column represents all movie ratings and no descriptive entity ratings, and the ``1/4'' column represents $1/4$'ths $3/4$'ths of all movie and descriptive entity ratings, respectively. 

%
For all models, when we see a statistical significant improvement in the \gls{ndcg} when adding descriptive entities, we also see the same statistical significant change when substituting ratings.
\emph{This indicates that descriptive entities carry more information than recommendable entities}.
%
In particular, Item \gls{knn} witnesses a $\sim 47.05\%$ increase in \gls{hr}, though this is primarily attributed the fact that the density of descriptive entity ratings is higher than that of recommendable entities.
In this setup, the best performing models are PPR-COLLAB, PPR-JOINT, and MF.
\emph{These results demonstrate the informational value of descriptive entity ratings, as we are now replacing data instead of simply adding more training data.} 
%
%
%
This would allow interviewing systems to focus on querying the user for opinions on more descriptive entities where the chance of getting a useful answer is higher. 
Yet, looking at the model performance in absolute values, we highlight how there is ample margin of improvement in the design of models that can exploit the informational value of non-recommendable entities in a knowledge graph.



%% file: sections/conclusions.tex

\section{Conclusion}\label{sec:conclusion}
We introduce MindReader, the first dataset with explicit user ratings on both recommendable and descriptive entities in a knowledge graph within the domain of movie recommendation.
We release both the dataset and the open-source platform for expanding the data-collection in other domains.
This will allow further research in KG-enhanced recommendation models.
Our analysis of the data shows that users are on average more likely to provide informative ratings towards descriptive entities compared to recommendable entities.
This observation is particularly important for interview-based cold-start recommender systems.
We also evaluate a variety of models for recommendation to assert the informational value of explicit descriptive entity ratings for recommendation. 
We find that including descriptive entity ratings in the training process increases the performance of almost all models thus justifying the potential of exploiting this kind of information.

%% file: sections/appendix.tex

\section{Exploration phase sampling}
We formalise the approach for entity sampling during the exploration phase in \autoref{alg:sampling}. For some entity $\entity$ in group $\adjacentgroup \subseteq \adjacents$, we sample it with probability \begin{equation}\label{eq:sample_exploration}
    \dfrac{S(\entity)}{\sum_{\entity\in \mathcal{A}_g}{S(e)}}
\end{equation} where $\globalpr:\entities \to \mathbb{R}$ is a function that returns the global PageRank of $e$ in $\graph$.

\begin{algorithm}
    \SetKwInOut{Input}{input}\SetKwInOut{Output}{output}
    \DontPrintSemicolon
    \SetNoFillComment
    
    \Input{A set $\adjacents$ of adjacent entities,\\ A number $n$ of entities to sample}
    \Output{A list of entities}
    \BlankLine
    
    $groups$ := Group entities in $\adjacents$ by type\;
    $groups$ := Shuffle $groups$ to ensure random order\;
    $results$ := Empty list\;
    \While{|results| < $n$ and $\exists \adjacentgroup \in groups:  |\adjacentgroup| > 0$}{
        \For{$R_g$ $\in$ $groups$}{
            Stop iterating if $|results| \geq n$\;
            $sample$ := Sample element from $\adjacentgroup$ by \autoref{eq:sample_exploration}\;
            Add $sample$ to $results$\;
            Remove $sample$ from $\adjacentgroup$\;
        }
    }
    
    \BlankLine
    \KwRet{$results$}
    \caption{Sampling of entities}\label{alg:sampling}
\end{algorithm}

\section{Analysis}

\begin{figure}[thb]
    \centering
    \includegraphics[width=\linewidth]{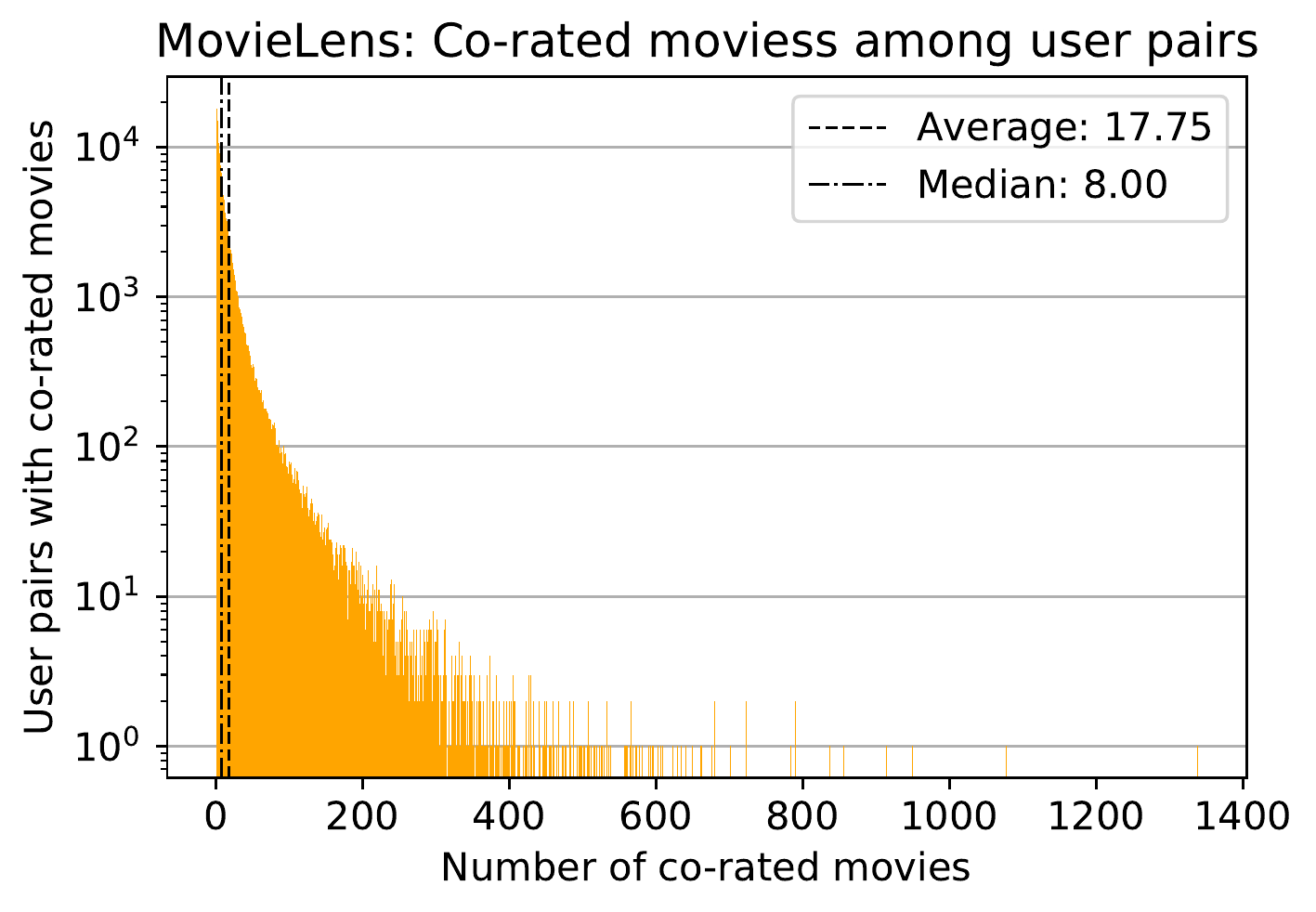}
    \Description{A figure illustrating how many \gls{ml} user pairs that have co-rated a number of movies. The x-axis is the number of co-ratings for a user pair, and the y-axis is many user pairs that share that amount of co-ratings.}
    \caption{A figure illustrating how many \gls{ml} user pairs that have co-rated a number of movies.}
    \label{fig:ml-co-ratings}
\end{figure}

\para{Co-ratings.} We observe that MovieLens has a higher number of co-ratings as seen in \autoref{fig:ml-co-ratings}. \gls{cf} models would therefore generally perform better on MovieLens than MindReader, as they have more information when comparing items and users.\\ 

\para{Label propagation.}\label{section:label_propagation}
We assume there is a correlation between descriptive and recommendable entities, meaning that having user preferences on one group gives on information about their preferences towards the other. As an initial test of this assumption, we implemented a label propagation approach as described in~\cite{zhu2002learning} on the \bin{} dataset using entities to predict descriptive entity ratings. We used a simple label propagation method, assigning $-1$ to disliked entities, $1$ to liked entities, and a default value of $0$ to all unrated entities. We propagated $10$ times for each user with no early stopping feature. The label propagation method assigns a score $\hat{o}_e\in\mathbb{R}$ to each entity $e$. The label was chosen using a threshold of $0$, i.e.  
\begin{equation*}
    \hat{l}_e = 
    \begin{cases}
    1         &   \text{if } \hat{o}_e \geq 0 \\
    -1  &   \text{otherwise}
    \end{cases}
\end{equation*}

converting $\hat{o}_e$ to the predicted label $\hat{l}_e$. Using the average number of descriptive entities presented in \autoref{tab:label-prop}, we can compute the like to dislike ratio to be $0.75$ being close to the ratio of MindReader, which is $0.73$. This indicates that the model predicts a realistic number of liked and disliked descriptive entities. We further used the MindReader ratio as a weight when randomly choosing a label for an entity, picking $1$ (like) with a probability of $0.73$, otherwise picking $-1$ (dislike). This the random method had an accuracy of $0.64$, while label propagation had an accuracy of $0.75$. This means that there is relationship between recommendable and descriptive entities in our \gls{kg}, as it performs better than a randomised approach on both likes and dislikes.

\begin{table*}[]
    \centering
    \begin{tabular}{l|r|r|r|c}
        \hline
        & \multicolumn{1}{l|}{\textbf{Precision}} & \multicolumn{1}{l|}{\textbf{Recall}} & \multicolumn{1}{l|}{\textbf{F1}} & \multicolumn{1}{l}{\textbf{Avg. No. DEs w/ label}} \\ \hline
        \textbf{Positive}      & 0.87 (0.81)            & 0.81 (0.72)            & 0.84 (0.76) & \multicolumn{1}{r}{14,240.7} \\ \hline
        \textbf{Negative}      & 0.37 (0.18)            & 0.50 (0.27)            & 0.43 (0.22) & \multicolumn{1}{r}{4,466.3}  \\ \hline
        \textbf{Accuracy}      & \multicolumn{1}{c|}{-} & \multicolumn{1}{c|}{-} & 0.75 (0.64) & -                             \\ \hline
        \textbf{Macro Avg.}    & 0.62 (0.50)            & 0.65 (0.49)            & 0.63 (0.49) & -                             \\ \hline
        \textbf{Weighted Avg.} & 0.78 (0.69)            & 0.75 (0.64)            & 0.76 (0.66) & -                             \\ \hline
        \end{tabular}
    \caption{The results of propagating movie labels to descriptive entities. The results using weighted random is shown in parentheses. The last column is the number of nodes the label propagation algorithm labelled as positive or negative. Furthermore, the macro shows the mean per label, while weighted includes the support into the mean.}
    \label{tab:label-prop}
\end{table*}



\para{Discriminative entities.}
Whether we are inferring preferences in a warm- or cold-start setting, the entities that a user likes and dislikes are both important parts of their preferences. Especially in a cold-start setting, it is equally as important to determine what a user likes as well as what they do not like as quickly as possible. While users are less likely to answer ``Don't know'' to a descriptive entity compared to a recommendable entity, if the users are mostly answering ``Like'', the information gained by asking towards descriptive entities may not be as significant as expected.

We investigate which entities in the MindReader dataset lead to most contention to identify which entities are the most effective discriminators in this regard.
For each rating category, we extract the entities that most commonly fall into the category. The most commonly occurring entities can be seen in \autoref{tab:top_10}.

\begin{table*}[h]
    \centering
    \begin{tabular}{|c|c|c|}
         \hline 
         \textbf{Like} & \textbf{Dislike} & \textbf{Don't know} \\\hline 
            Movies of the 2000s  & Romantic Comedy  & Movies of the 1980s \\
            (805) & (182) & (140) \\\hline

            Movies of the 2010s  & \textbf{Action Film}  & Warner Bros \\
            (756) & (136) & (129) \\\hline

            \textbf{Drama Film}  & Superhero Film  & Columbia Pictures \\
            (736) & (110) & (120) \\\hline

            Movies of the 1990s  & Horror Film  & Movies of the 1990s \\
            (665) & (101) & (119) \\\hline

            \textbf{Comedy Film}  & Romance Film  & 20th Century Fox \\
            (652) & (98) & (117) \\\hline

            \textbf{Action Film}  & Movies of the 1980s  & Universal Studios \\
            (607) & (80) & (110) \\\hline

            Science Fiction Film  & Spy Film  & Paramount Pictures \\
            (523) & (66) & (110) \\\hline

            Film Based On Literature  & \textbf{Comedy Film}  & Movies of the 2010s \\
            (505) & (64) & (106) \\\hline

            Fantasy Film  & \textbf{Drama Film}  & Speculative Fiction Film \\
            (472) & (64) & (97) \\\hline

            Adventure Film  & Musical Film  & Crime Film \\
            (394) & (61) & (95) \\\hline
         
    \end{tabular}
    \caption{Top 10 entities that most commonly appeared in each answer category. For each entity, the number of observations for the entity in that category is shown below its name. Entities present in both the top liked and disliked lists are marked with bold.}
    \label{tab:top_10}
\end{table*}

From the results in \autoref{tab:top_10}, we see that some descriptive entities are effective discriminators of user opinions. The descriptive entities \textbf{Drama Film}, \textbf{Comedy Film}, and \textbf{Action Film} all occur both in the top liked and disliked entities. Of course, their high ranking in this list is of little surprise as we have already seen that users are more likely to answer ``Like'' or ``Dislike'' when asked towards a descriptive entity. However, these results suggest that not only are users familiar with these entities, but the users' opinions on the entities are not one-sided.

\section{Results}
This section contains result tables for some of the referenced experiments. 
The results from experiment \textbf{(a)} and \textbf{(b)} with the most popular movies included in the test set are shown in \autoref{tab:full_results_wtp}. All models, barring TransE, PPR-JOINT and PPR-COLLAB, either maintain comparable performance or witness a statistically significant increase when provided with descriptive entity ratings, both when simply including the ratings (experiment \textbf{(a)}) and replacing movie ratings (experiment \textbf{(b)}). As PPR-JOINT and PPR-COLLAB inherently model TopPop, we assume they decrease because there is less knowledge about popular entities. \\

The results from experiment \textbf{(c)} are shown in \autoref{table:wtp-substituting_no_movies_hr} and \autoref{table:wtp-substituting_no_movies_ndcg}. All models, except TransH, witness a statistically significant decrease in performance when removing movie ratings without replacement, while TransH maintains performance.

This may be an indication that hybrid methods can benefit greatly from descriptive entity ratings. 
However, although we did not use \gls{ppr} to provide recommendations for the MindReader users, some of the movies asked about were sampled from the global PageRank of nodes in the \gls{kg}. Nevertheless, the results can be explained partially by how (personalised) PageRank generates recommendations. 
In the PPR-COLLAB model, the ranks of entities should capture their popularity, as entities with many in-going edges from users will be determined to be central to the graph, and thereby be highly ranked, achieving very similar performance to TopPop. 
With the addition of edges between related entities as in PPR-JOINT, an entity being highly ranked due to its popularity will also increase the rank of its related entities, as it effectively serves as a proxy for its related entities from the point of view of the random surfer modelled in PageRank. 

The translational Trans* models, when including the most popular movies in the test set, see no statistically significant change in performance except for TransH, which sees an improvement when adding descriptive entity ratings to the training data.
The translational TransH models see a statistical significant increase in performance, while the TransE models decrease, possibly due to the fact the that the TransH models are more expressive and better capture relations than TransE models.
Generally, we see that the TransE models outperform their hyperplane-based counterparts. 
While TransH is designed to afford greater expressiveness of embeddings, the limitations of TransE embeddings may be beneficial in modelling entity popularity. When many users like the same entity, optimising the TransE objective in~\cite{TransE} requires all users to be embedded similarly. 
This forces the embeddings to be driven primarily by the number of ratings, making all users' embeddings most similar to the entity with the most ratings; the most popular entity.
This theory is supported by the results shown in~\autoref{tab:full_results_wtp} vs. \autoref{tab:full_results}, where the performance of the TransE models drop as we exclude the most popular movies from the test set.

We expect that this popularity effect is also partially the cause of TransH performing worse than TransE, as the popularity effect has a lesser impact on the embeddings of entities as these are now distributed across relations. 
For the same reason, TransH-KG is able to reach comparable performance to that of TransE and TransE-KG when provided with \gls{kg} triples as part of the training data, and is further able to outperform both TransE and TransE-KG once the most popular movies are excluded from the test set. 

Finally, considering all the embedding-based models (barring TransE due to its limited expressiveness), the performance is increased when provided with descriptive entity ratings. The reason behind this may be related to the work of~\cite{MutliTaskingCrossCompress}, which demonstrates a positive correlation between the number of co-ratings and shared neighbours between entity pairs in a \gls{kg}. 
As all the embedding models operate on a notion of user-entity embedding similarity, it can be argued that the models are able to perform better as the user embeddings are learned to be close to the descriptive entities which, in turn, are learned to be close to their related recommendable entities in the latent space.

Statistical significance is derived from a paired sample $t$-test between $4/4$ and $n/4$ distribution pairs when $n < 4$.

The purpose of these experiments is to determine if and how well descriptive entity ratings can serve as replacements for recommendable entity ratings in recommender systems.
, which is one of our research questions posed in~\autoref{sec:introduction}. 

For most models, we observe that the change in performance when substituting with descriptive entity ratings is, perhaps surprisingly, either statistically insignificant or a statistically significant improvement, both in terms of \gls{hr}$@k$ and \gls{ndcg}$@k$. 
Exceptions to this are the TopPop and TransE models, for which we observe a statistically significant decrease in performance.

as seen in \autoref{table:related}. 

As in the previous experiments, we observe that TopPop performs well in these experiments. However, as we randomly replace movie ratings with descriptive entity ratings, we have a higher probability of replacing ratings on popular movies as they occur more frequently in the training set. This is why we see a statistically significant decrease in the performance of TopPop in \autoref{tab:full_results} and \autoref{tab:full_results_wtp} even when we do not exclude top popular movies from the test sets. Similarly, the learning models are less prone to predict top popular movies which generally leads to reduced performance compared to the previous experiments.

As shown in \autoref{fig:rating_distributions_over_labels}, a user is generally more likely to provide a like or dislike rating on a given descriptive entity than a recommendable entity. 
Since users are more likely to be familiar with descriptive entities, being able to achieve comparable recommendation quality given more descriptive entity ratings than recommendable entity ratings is desirable especially for interview-based cold-start recommendations. 
In such systems, it is critical to query the user for questions that the user is likely to provide useful answers to. 


\para{Removing movie ratings.}\label{sec:removing_movies}
The results of experiment \textbf{(c)} are shown in \autoref{table:wtp-substituting_no_movies_hr}, \autoref{table:wtp-substituting_no_movies_ndcg}, \autoref{table:ntp_substituting_no_movies_hr}, and \autoref{table:ntp_substituting_no_movies_ndcg}. We find that all models deteriorate in performance with statistical significance when movie ratings are removed without being substituted by descriptive entity ratings. While this result may be of little surprise, it shows that the substituting descriptive entity ratings in experiment \textbf{(b)} provide the models with information that generally affords similar quality of recommendations.

\input{Tables/wtp-substituting-hr.tex}
\FloatBarrier
\input{Tables/wtp-remove-only-hr.tex}
\input{Tables/wtp-remove-only-ndcg.tex}
\FloatBarrier
\input{Tables/ntp-remove-only-hr.tex}
\input{Tables/ntp-remove-only-ndcg.tex}

%% file: Tables/wtp-substituting-hr.tex
\begin{table*}[ht!]
        \centering
        \begin{adjustbox}{max width=\linewidth}
        \begin{tabular}{|l|c c|c c c c c|c c|c c c c c|}
                    \hline
                    & \multicolumn{7}{|c|}{\textbf{HR@10}} & \multicolumn{7}{|c|}{\textbf{NDCG@10}}\\\hline
                    Models & All movies & All entities & & 4/4 & 3/4 & 2/4 & 1/4 & All movies & All entities & & 4/4 & 3/4 & 2/4 & 1/4
                    \\\hline
                     BPR & $0.50 \pm 0.10$ & $0.58 \pm 0.02$ & & $0.54 \pm 0.01$ & $0.55 \pm 0.02$ & $0.52 \pm 0.01^*$ & $0.54 \pm 0.02$ & $0.34 \pm 0.07$ & $0.39 \pm 0.01^*$ & & $0.35 \pm 0.01$ & $0.37 \pm 0.01$ & $0.33 \pm 0.01^*$ & $0.35 \pm 0.01$
                    \\ MF & $0.57 \pm 0.02$ & $0.61 \pm 0.02^*$ & & $0.57 \pm 0.02$ & $\mathbf{0.58 \pm 0.02}$ & $\mathbf{0.56 \pm 0.01}$ & $0.57 \pm 0.01$ & $\mathbf{0.39 \pm 0.01}$ & $0.42 \pm 0.01^*$ & & $\mathbf{0.39 \pm 0.01}$ & $\mathbf{0.39 \pm 0.01}$ & $\mathbf{0.38 \pm 0.01^*}$ & $0.37 \pm 0.01^*$
                    \\ Item kNN & $0.23 \pm 0.02$ & $0.51 \pm 0.02^*$ & & $0.23 \pm 0.01$ & $0.26 \pm 0.01^*$ & $0.29 \pm 0.01^*$ & $0.34 \pm 0.02^*$ & $0.13 \pm 0.01$ & $0.36 \pm 0.01^*$ & & $0.13 \pm 0.01$ & $0.15 \pm 0.01^*$ & $0.18 \pm 0.01^*$ & $0.22 \pm 0.01^*$
                    \\ User kNN & $0.44 \pm 0.02$ & $0.56 \pm 0.02^*$ & & $0.45 \pm 0.01$ & $0.45 \pm 0.02$ & $0.46 \pm 0.01$ & $0.46 \pm 0.02$ & $0.31 \pm 0.01$ & $0.39 \pm 0.01^*$ & & $0.32 \pm 0.01$ & $0.31 \pm 0.01$ & $0.32 \pm 0.01$ & $0.31 \pm 0.01$
                    \\ PPR-COLLAB & $\mathbf{0.58 \pm 0.01}$ & $0.61 \pm 0.02^*$ & & $\mathbf{0.58 \pm 0.01}$ & $0.56 \pm 0.02$ & $\mathbf{0.56 \pm 0.01^*}$ & $\mathbf{0.58 \pm 0.02}$ & $\mathbf{0.39 \pm 0.01}$ & $0.41 \pm 0.01^*$ & & $\mathbf{0.39 \pm 0.01}$ & $0.38 \pm 0.02$ & $0.37 \pm 0.01^*$ & $\mathbf{0.38 \pm 0.01^*}$
                    \\ PPR-JOINT & $0.54 \pm 0.01$ & $\mathbf{0.64 \pm 0.01^*}$ & & $0.54 \pm 0.01$ & $0.52 \pm 0.01^*$ & $0.51 \pm 0.02^*$ & $0.53 \pm 0.01^*$ & $0.36 \pm 0.01$ & $\mathbf{0.46 \pm 0.01^*}$ & & $0.37 \pm 0.01$ & $0.35 \pm 0.01^*$ & $0.35 \pm 0.01^*$ & $0.34 \pm 0.01^*$
                    \\ PPR-KG & $0.33 \pm 0.01$ & $0.52 \pm 0.01^*$ & & $0.33 \pm 0.02$ & $0.36 \pm 0.01^*$ & $0.39 \pm 0.02^*$ & $0.44 \pm 0.02^*$ & $0.20 \pm 0.01$ & $0.37 \pm 0.01^*$ & & $0.20 \pm 0.01$ & $0.23 \pm 0.00^*$ & $0.26 \pm 0.01^*$ & $0.30 \pm 0.01^*$
                    \\ TopPop & $\mathbf{0.58 \pm 0.01}$ & $0.58 \pm 0.02$ & & $\mathbf{0.58 \pm 0.02}$ & $0.56 \pm 0.02$ & $\mathbf{0.56 \pm 0.01^*}$ & $0.56 \pm 0.02^*$ & $\mathbf{0.39 \pm 0.01}$ & $0.39 \pm 0.00$ & & $\mathbf{0.39 \pm 0.01}$ & $0.38 \pm 0.01$ & $0.37 \pm 0.01^*$ & $0.37 \pm 0.01^*$
                    \\ TransE & $0.45 \pm 0.06$ & $0.46 \pm 0.08$ & & $0.52 \pm 0.01$ & $0.49 \pm 0.03^*$ & $0.46 \pm 0.01^*$ & $0.44 \pm 0.02^*$ & $0.28 \pm 0.05$ & $0.29 \pm 0.07$ & & $0.34 \pm 0.01$ & $0.31 \pm 0.02^*$ & $0.28 \pm 0.01^*$ & $0.27 \pm 0.02^*$
                    \\ TransE-KG & $0.45 \pm 0.06$ & $0.39 \pm 0.09$ & & $0.54 \pm 0.01$ & $0.47 \pm 0.02^*$ & $0.43 \pm 0.02^*$ & $0.33 \pm 0.02^*$ & $0.28 \pm 0.04$ & $0.25 \pm 0.06$ & & $0.34 \pm 0.01$ & $0.31 \pm 0.01^*$ & $0.28 \pm 0.01^*$ & $0.20 \pm 0.02^*$
                    \\ TransH & $0.34 \pm 0.06$ & $0.46 \pm 0.04^*$ & & $0.36 \pm 0.09$ & $0.38 \pm 0.06$ & $0.36 \pm 0.04$ & $0.33 \pm 0.05$ & $0.21 \pm 0.04$ & $0.30 \pm 0.04^*$ & & $0.23 \pm 0.06$ & $0.24 \pm 0.04$ & $0.22 \pm 0.03$ & $0.20 \pm 0.04$
                    \\ TransH-KG & $0.44 \pm 0.09$ & $0.48 \pm 0.04$ & & $0.39 \pm 0.08$ & $0.47 \pm 0.07^*$ & $0.46 \pm 0.03^*$ & $0.41 \pm 0.05$ & $0.28 \pm 0.07$ & $0.31 \pm 0.03$ & & $0.24 \pm 0.06$ & $0.30 \pm 0.05^*$ & $0.30 \pm 0.03^*$ & $0.25 \pm 0.03$
                    \\\hline
            \end{tabular}
        \end{adjustbox}
        \caption{\{HR, NDCG\}@10 performance including popular movies. Statistically significant differences in mean performance between related experiment pairs are marked with a star (\textbf{*}).  }\label{tab:full_results_wtp}
\end{table*}

%% file: Tables/wtp-remove-only-hr.tex
\begin{table}[!ht]
        \centering
        \begin{adjustbox}{max width=\linewidth}
        \begin{tabular}{l|c|c}
                \hline
                \multicolumn{1}{c|}{Models} & 4/4 & 1/4 (no DEs)
                \\\hline
                 BPR & $0.54 \pm 0.01$ & $0.29 \pm 0.01^*$
                \\ Item kNN & $0.23 \pm 0.01$ & $0.09 \pm 0.01^*$
                \\ MF & $0.57 \pm 0.02$ & $0.41 \pm 0.02^*$
                \\ PPR-COLLAB & $\mathbf{0.58 \pm 0.01}$ & $0.42 \pm 0.01^*$
                \\ PPR-JOINT & $0.54 \pm 0.01$ & $0.31 \pm 0.02^*$
                \\ PPR-KG & $0.33 \pm 0.02$ & $0.24 \pm 0.02^*$
                \\ TopPop & $\mathbf{0.58 \pm 0.02}$ & $\mathbf{0.55 \pm 0.01^*}$
                \\ TransE & $0.52 \pm 0.01$ & $0.37 \pm 0.01^*$
                \\ TransE-KG & $0.54 \pm 0.01$ & $0.35 \pm 0.01^*$
                \\ TransH & $0.36 \pm 0.09$ & $0.37 \pm 0.12$
                \\ TransH-KG & $0.39 \pm 0.08$ & $0.22 \pm 0.10^*$
                \\ User kNN & $0.45 \pm 0.01$ & $0.10 \pm 0.01^*$
                \\\hline
        \end{tabular}
        \end{adjustbox}
        \caption{\gls{hr}$@10$ performance including popular movies where removed movie ratings are not substituted by descriptive entity ratings.}
        \label{table:wtp-substituting_no_movies_hr}
\end{table}

%% file: Tables/wtp-remove-only-ndcg.tex
\begin{table}[!ht]
        \centering
        \begin{adjustbox}{max width=\linewidth}
        \begin{tabular}{l|c|c}
                \hline
                \multicolumn{1}{c|}{Models} & 4/4 & 1/4 (no DEs)
                \\\hline
                 BPR & $0.35 \pm 0.01$ & $0.19 \pm 0.01^*$
                \\ Item kNN & $0.13 \pm 0.01$ & $0.05 \pm 0.01^*$
                \\ MF & $\mathbf{0.39 \pm 0.01}$ & $0.27 \pm 0.02^*$
                \\ PPR-COLLAB & $\mathbf{0.39 \pm 0.01}$ & $0.24 \pm 0.01^*$
                \\ PPR-JOINT & $0.37 \pm 0.01$ & $0.17 \pm 0.01^*$
                \\ PPR-KG & $0.20 \pm 0.01$ & $0.14 \pm 0.01^*$
                \\ TopPop & $\mathbf{0.39 \pm 0.01}$ & $\mathbf{0.36 \pm 0.01^*}$
                \\ TransE & $0.34 \pm 0.01$ & $0.24 \pm 0.01^*$
                \\ TransE-KG & $0.34 \pm 0.01$ & $0.22 \pm 0.01^*$
                \\ TransH & $0.23 \pm 0.06$ & $0.24 \pm 0.09$
                \\ TransH-KG & $0.24 \pm 0.06$ & $0.13 \pm 0.07^*$
                \\ User kNN & $0.32 \pm 0.01$ & $0.07 \pm 0.01^*$
                \\\hline
        \end{tabular}
        \end{adjustbox}
        \caption{\gls{ndcg}$@10$ performance including popular movies where removed movie ratings are not substituted by descriptive entity ratings.}
        \label{table:wtp-substituting_no_movies_ndcg}
\end{table}

%% file: Tables/ntp-remove-only-hr.tex
\begin{table}[ht!]
	\centering
	\begin{adjustbox}{max width=\linewidth}
	\begin{tabular}{l|c|c}
		\hline
		\multicolumn{1}{c|}{Models} & All movies & 1/4 (no DEs)
		\\\hline
		 BPR & $0.36 \pm 0.06$ & $0.24 \pm 0.02^*$
		\\ Item kNN & $0.17 \pm 0.01$ & $0.08 \pm 0.01^*$
		\\ MF & $0.42 \pm 0.01$ & $0.25 \pm 0.02^*$
		\\ PPR-COLLAB & $0.42 \pm 0.01$ & $0.28 \pm 0.02^*$
		\\ PPR-JOINT & $0.38 \pm 0.01$ & $0.23 \pm 0.02^*$
		\\ PPR-KG & $0.25 \pm 0.01$ & $0.21 \pm 0.01^*$
		\\ TopPop & $\mathbf{0.43 \pm 0.01}$ & $\mathbf{0.42 \pm 0.01}$
		\\ TransE & $0.33 \pm 0.03$ & $0.21 \pm 0.02^*$
		\\ TransE-KG & $0.28 \pm 0.01$ & $0.21 \pm 0.02^*$
		\\ TransH & $0.28 \pm 0.04$ & $0.32 \pm 0.01^*$
		\\ TransH-KG & $0.30 \pm 0.04$ & $0.23 \pm 0.01^*$
		\\ User kNN & $0.31 \pm 0.02$ & $0.08 \pm 0.01^*$
		\\\hline
	\end{tabular}
	\end{adjustbox}
	\caption{\gls{hr}$@10$ performance without popular movies where removed movie ratings are not substituted by descriptive entity ratings.}
        \label{table:ntp_substituting_no_movies_hr}
\end{table}

%% file: Tables/ntp-remove-only-ndcg.tex
\begin{table}[ht!]
	\centering
	\begin{adjustbox}{max width=\linewidth}
	\begin{tabular}{l|c|c}
		\hline
		\multicolumn{1}{c|}{Models} & All movies & 1/4 (no DEs)
		\\\hline
		 BPR & $0.18 \pm 0.02$ & $0.14 \pm 0.01^*$
		\\ Item kNN & $0.09 \pm 0.01$ & $0.05 \pm 0.01^*$
		\\ MF & $\mathbf{0.20 \pm 0.01}$ & $0.13 \pm 0.01^*$
		\\ PPR-COLLAB & $\mathbf{0.20 \pm 0.01}$ & $0.14 \pm 0.01^*$
		\\ PPR-JOINT & $0.19 \pm 0.01$ & $0.13 \pm 0.01^*$
		\\ PPR-KG & $0.15 \pm 0.01$ & $0.12 \pm 0.01^*$
		\\ TopPop & $\mathbf{0.20 \pm 0.01}$ & $\mathbf{0.19 \pm 0.01}$
		\\ TransE & $0.18 \pm 0.01$ & $0.11 \pm 0.01^*$
		\\ TransE-KG & $0.15 \pm 0.01$ & $0.12 \pm 0.01^*$
		\\ TransH & $0.15 \pm 0.02$ & $0.17 \pm 0.01$
		\\ TransH-KG & $0.16 \pm 0.02$ & $0.12 \pm 0.01^*$
		\\ User kNN & $0.17 \pm 0.01$ & $0.05 \pm 0.01^*$
		\\\hline
	\end{tabular}
	\end{adjustbox}
	\caption{\gls{ndcg}$@10$ performance without popular movies where removed movie ratings are not substituted by descriptive entity ratings.}
        \label{table:ntp_substituting_no_movies_ndcg}
\end{table}